\documentclass[%
 aip,
 jmp,%
 amsmath,amssymb,
 reprint,%
]{revtex4-1}

\usepackage[T1]{fontenc}
\usepackage[utf8]{inputenc}
\usepackage{amsmath}
\usepackage{amssymb}
\usepackage{graphicx}
\usepackage{refstyle}

\usepackage[
  citecolor=blue,
  colorlinks,
  linkcolor=blue,
  urlcolor=blue,
]{hyperref}
\usepackage[dvipsnames]{xcolor}
\usepackage{dcolumn}
\usepackage{bm}

\begin{document}

\preprint{AIP/123-QED}

\title{{Effect of chaotic agent dynamics on coevolution of cooperation and synchronization}}
\author{Rohitashwa Chattopadhyay}
\email{crohit@iitk.ac.in}
\affiliation{
  Department of Physics,
  Indian Institute of Technology Kanpur,
  Uttar Pradesh 208016, India
}
\author{Shubhadeep Sadhukhan}
\email{deep@iitk.ac.in}
\affiliation{
  Department of Physics,
  Indian Institute of Technology Kanpur,
  Uttar Pradesh 208016, India
}

\author{Sagar Chakraborty}
\email{sagarc@iitk.ac.in}
\affiliation{
  Department of Physics,
  Indian Institute of Technology Kanpur,
  Uttar Pradesh 208016, India
}

\date{\today}

\begin{abstract}
The effect of the chaotic dynamical states of the agents on the coevolution of  cooperation and synchronization in a structured population of the agents remains unexplored. With a view to gaining insights into this problem, we construct a coupled map lattice of the paradigmatic chaotic logistic map by adopting the Watts--Strogatz network algorithm. The map models the agent's chaotic state dynamics. In the model, an agent benefits by synchronizing with its neighbours and in the process of doing so, it pays a cost. The agents update their strategies (cooperation or defection) by using either a stochastic or a deterministic rule in an attempt to fetch themselves higher payoffs than what they already have. Among some other interesting results, we find that beyond a critical coupling strength, that increases with the rewiring probability parameter of the Watts--Strogatz model, the coupled map lattice is spatiotemporally synchronized regardless of the rewiring probability. Moreover, we observe that the population does not desynchronize completely---and hence finite level of cooperation is sustained---even when the average degree of the coupled map lattice is very high. These results are at odds with how a population of the non-chaotic Kuramoto oscillators as agents would behave. Our model also brings forth the possibility of the emergence of cooperation through synchronization onto a dynamical state that is a periodic orbit attractor.
\end{abstract}
\maketitle
\section{Introduction}
\label{sec:I}

Cooperation~\cite{1998_Smith, 2013_NC, 2011_Bourke} plays a pivotal role in the sustenance of the biological, the social, and the economic systems. An insightful theoretical approach for studying the emergence of cooperation in such systems is the evolutionary game theory~\cite{1982_Smith, 2006_Axelrod, 2006_Nowak,2009_Gintis,2017_PJRWS_PR}. In the framework of the theory, each individual or agent in the population adopts a strategy that is based on the payoff it receives by interacting with the other agents. It is generally observed that an individual's self-interest acts as an obstacle in the emergence of cooperation, e.g., in the stylized game of the prisoner's dilemma~\cite{1965_RC}. Thus, one of the main objectives in the theory is to understand the situations where even though each agent has a higher incentive to defect, how the global cooperation-state emerges. To this end, many models and mechanisms for the emergence of global cooperation are available in the research literature~\cite{1992_Nowak,2006_Nowak,2006_OHLN_Nature,2010_PS_B,2011_WCG,2011_RH_PNAS,2018_Hilbe, 2020_MMC_PRE}. \textcolor{black}{The coevolutionary processes---the processes that co-occur with the evolution of strategies---are known to promote cooperation effectively~\cite{2009_SP_EPL,2010_PS_B}}. Also, certain types of interaction between the agents affect the degree of cooperation present in the system~\cite{2016_AWPSJ_PRE,2017_BPL_NJP,2019_L_EPL}.

The individual agents may have some associated dynamics---e.g., replicator dynamics~\cite{1978_TJ_MB,2002_SAF_PNAS}, best response dynamics~\cite{1992_M_JET}, imitation dynamics~\cite{1994_H_JMS}, sampling dynamics~\cite{2000_S_GEB}, and opinion dynamics~\cite{2009_CFL_RMP}---modelling the evolution of their states. In a networked population of agents (who can either cooperate or defect) when arranged so as to have the Kuramoto model~\cite{1975_Kuramoto} realized as far as their coupled dynamics is concerned, the costly interactions present a dilemma, termed the evolutionary Kuramoto dilemma~\cite{2017_Antonioni_Cardillo_PRL}: An agent may cooperate by paying the cost in order to have its state synchronized with the rest agents in the network; or it may defect and thus not suffer any cost, while expecting the other agents' states to be synchronized to its own. Just as the study of emergence of cooperation has a long history in the evolutionary game dynamics, the emergence of synchronization in complex nonlinear systems has its own very rich literature~\cite{2005_ABVCR_RMP, 2006_BLMCH_PR, 2008_AGKMZ_PR}. The interplay of synchonization and cooperation in the evolutionary game dynamics of the agents is dependent on the topological details of the network employed. For example, in Watts--Strogatz (WS) network~\cite{1998_WS_Nature} of the aforementioned agents, the population of reaches fully synchronization state only for  high values rewiring probability and the coupling strength beyond which such a state is attained, decreases with an increase in the random rewiring probability~\cite{2018_Liu_Wu_Guan_EPL}. Furthermore, in the WS network, and also in the Barab\'{a}si--Albert network~\cite{1999_BA_S} and the Erd\H{o}s--R\'{e}nyi network~\cite{1959_ER_PMD}, the increase in average degree of the nodes, desynchronizes the agents' dynamics~\cite{2018_Yang_Han-Xin_Zhou}. 

{The primary question we ask in this paper is, given a network topology, how robust such conclusions regarding the behaviour of the population of the agents are if the uncoupled agent dynamics is chaotic.} In this context, we recall that in the Kuramoto model of the population of the agents, the uncoupled agent dynamics is that of a uniform oscillator on a circle, i.e., it is a flow on a 1-torus such that the phase uniformly changes. The Kuramoto model is synonymous with the homogeneous sinusoidal coupling between such non-chaotic oscillators, and once endowed with such coupling, the oscillators may now be termed as the Kuramoto oscillators. It is all too well known that globally coupled identical Kuramoto oscillators can not show chaotic dynamics~\cite{1993_WS_PRL} and any chaos that may be there in the  globally coupled non-identical Kuramoto oscillators vanishes in the infinite size limit of the population. However, it may be possible to have chaotic state of the population in the thermodynamic limit if the condition of global coupling between the non-identical Kuramoto oscillators is done away with~\cite{2018_BPM_C}. Our quest, however, seeks a non-chaotic synchronized population state when the individual uncoupled agent dynamics are identical and chaotic. \textcolor{black}{ More specifically, we are interested in the correlation between the co-occurrence of the temporal changes in the cooperator and the synchronization levels of such a population; or in other words, in the co-evolution of the cooperation and the synchronization.}

Since, discrete dynamical systems---often called maps---are capable of showing chaotic behaviour even with single phase space variable, in this paper it conveniently suits our purpose to work with the nonlinear maps. Specifically, we work with the paradigmatic chaotic logistic map~\cite{1976_May_N} as a toy model. Another reason behind using the logistic map is that its phase variable, when scaled by a constant factor of $2\pi$, is bounded between zero to $2\pi$ just like the phase of the Kuramoto oscillator. One extensively studied model for a large number of coupled maps is that of the coupled map lattice~(CML)~\cite{2014_Kaneko}, that was introduced as a simple tool to study the chaotic behaviour of the spatially extended systems. A CML consists of a lattice of maps, each map being on a unique lattice point that is coupled locally to other lattice points via the edges of the lattice. Despite being a rather simple construction, the CML has found extensive applications in modelling a broad spectrum of systems~\cite{1993_Kaneko}, e.g., pattern formation, crystal growth, the Josephson junction arrays, multi-mode lasers, and vortex dynamics. 

The synchronization of the coupled chaotic maps on different network topology and under various settings is also an intensively studied topic~\cite{2001_JJ_PRE,2003_MM_PRE,2004_LGH_PRE,2005_JAH_PRE,2005_AJH_PRE,2008_PMM_EPJB,2009_LXSXK_Chaos}. \textcolor{black}{It should be pointed out that there is a difference between the synchronization on to a fixed point in a CML and the synchronization among the Kuramoto oscillators: In the case of the CML, the synchronization on to a fixed point implies that all the lattice points have the identical non-oscillatory static state, while the synchronization in a network of the Kuramoto oscillators implies a dynamic state where all the synchronized oscillators' states change in unison with time. The former phenomenon has a close resemblance to the phenomenon of the amplitude death in the coupled limit cycle oscillators like the coupled Stuart Landau oscillators~\cite{1990_MS,1998_RSJ}.}
	
As already mentioned, since our aim is to see the effect of chaos in a given network, in this paper we choose to work with the WS model of generating networks that can be interpolations between a regular network and (almost) the Erd\H{o}s--R\'{e}nyi network depending on the values of the random rewiring probability. What we find is that the conclusions obtained for the behaviour of cooperators and defectors in a network vis-a-vis the evolutionary Kuramoto dilemma is so strongly dependent of the uncoupled agent's dynamics that they are in stark contrast with what is known to happen when rewiring probability or average degree changes in the Kuramoto model. Without further ado, in what follows, we introduce our model in Sec.~\ref{sec:model} and the results in Sec~\ref{sec:R}, before concluding in Sec.~\ref{sec:DnC}. 
\section{Model}
\label{sec:model} 
We consider a population of $N$ agents, each of which has its state defined by $x_n^i$ ($n$ being the time step and $i$ the index for the node) that evolves in accordance with the chaotic logistic map---$x^i_{n+1}= 4x^i_n\left(1-x^i_n\right)$---when the agents are not interacting with each other. Here, $\,i\in\{1,2,\cdots,N\}$, $n\in\{0,1,2,\cdots,\}$, and by construction, $x_n^i\in[0,1]\,\forall i,n$. As mentioned earlier, this is one of the simplest and most well-studied map that possesses chaotic solutions. The corresponding CML is defined mathematically as,
\begin{eqnarray}
	x^i_{n+1}=4 x^i_n\left(1-x^i_n\right)(1-s_n^i\epsilon)+\frac{s_n^i\epsilon}{k_i}\sum_{j=1}^{N} a_{ij}x_n^{j},
	\label{eq_rewired_strategy_network}
\end{eqnarray}
where $k_i$ is the degree of $i$th lattice site or node, $\epsilon\in[0,1]$ denotes the coupling constant quantifying the strength of the coupling, and $a_{ij}\in\{0,1\}$ are the elements of the adjacency matrix. This model effectively associates a strategy, $s_n^i$, to each agent at time $n$; $s_n^i$ can either be zero (defection) or unity (cooperation). Note that we are considering a simple but non-trivial linear coupling between the agents~\cite{2002_Sinha_PRE} so that $x_n^i\in[0,1]\,\forall i,n$ even when the agents interact. 

Now the idea is that a cooperator (an agent with cooperation strategy) chooses to interact with the other agents (as allowed in accordance with the topology of the CML) and endeavours to synchronize with their dynamics while incurring the cost of the interactions; whereas the defectors (agents with defection strategy) do not interact at all and hence incur no cost. The cost associated with a cooperating agent is the measure of rate of deviation from its underlying dynamics. Thus, the cost, $c_{n}^i$, for the $i$th agent at the $n${th} time step may be defined as as follows:
\begin{eqnarray}
	\label{eq:cost}
	&&c_{n}^i:=\alpha\left|\left\{x^i_{n}-4 x^i_{n-1}(1-x^i_{n-1})\right\}\right. \nonumber\\
	&& \phantom{c_{n}^i:=|}-\left.\left\{x^i_{n-1}+4 x^i_{n-2}(1-x^i_{n-2})\right\}\right|.
\end{eqnarray}
where, $\alpha$ is a positive real parameter termed the relative cost. One may note that this is exactly in line with the cost defined in the analogous Kuramoto model~\cite{2017_Antonioni_Cardillo_PRL}. The benefit, $b^i_n$ that an agent reaps is measured through the extent to which the agent synchronizes with its neighbours. We define it as
\begin{equation}
	\label{eq:benefit}
	b_n^i:=\frac{\sum_j r_n^{ij} a_{ij}}{\sum_j a_{ij}},
\end{equation}
where
\begin{equation}
	r_n^{ij}:=\bigg |\frac{\exp\left(2\pi \sqrt{-1}x_n^i\right)+\exp\left(2\pi \sqrt{-1} x_n^j\right)}{2}\bigg|,
\end{equation}  
quantifies for the degree or synchronization between a pair of agents, say, $i$th and $j$th. Note that $r_n^{ij}$ is unity (maximum) when $x_n^i=x_n^j$. We realize that $b_n^i$ is the local version of the global synchronization parameter~\cite{2016_Wang}, $r_G$, defined as,
\begin{equation}
	\label{Global_order_parameter}
	r_G :=\frac{|\sum_{i=1}^N \exp\left(2\pi \sqrt{-1} x^i\right)|}{N}.
\end{equation}
The system is said to be completely synchronized when $r_G=1$ that corresponds to the population state where $x^i_n=x^j_n$ $\forall i,j\in\{1,2,\cdots,N\}$. For smaller values of $r_G$, the population is away from the synchronized state; $r_G=0$ corresponds to fully unsynchronized or incoherent state that is realized as a random distribution of the individual agents' states. 

Having defined the cost and the benefit, we can now define the payoff \textcolor{black}{(or utility)} of $i${th} agent at the $n${th} time step for this interaction as \textcolor{black}{$U_n^i:=b_n^i-c_n^i$}. In our model, we include a rule for strategy update for each agent at each point of time; the $i$th agent would choose between cooperation ($s^i_n=1$) or defection ($s^i_n=0$) at time step $n$ depending on how it fared against the agents it interacted with at the time step $n$. This rule can be implemented either deterministically or stochastically. In deterministic strategy update, also known as the unconditional imitation rule, at each time step the agents at each node compare their own payoffs with all their neighbours and adopt the strategy of the most successful agent (with highest payoff) among them (the agent and its neighbours). A popular stochastic strategy exchange rule is the Fermi rule for strategy exchange~\cite{1993_B_GEB,1998_SHH_PRE,2009_RCS_PLR}. In this rule, at time $n$ with probability $\textcolor{black}{[1+\exp\{-\beta(U_{n}^{j}-U_{n}^{i})\}]^{-1}}$, the $i$th agent chooses the strategy of one of the randomly chosen agent, say $j$th agent, among all the agents it interacted with. Here, $\beta$ denotes the rationality factor. 

It is quite obvious that under such update rules, an initial state of the population such that all agents are cooperators does not change with time; same is true for the initial state where all the agents are defectors. An interesting observation concerning the difference between these two invariant states is that while the former state may or may not simultaneously be a synchronized state, the latter can never be a synchronized state as it is accompanied with no coupling between the agents. It is also interesting to note that sometimes while a completely coupled set of chaotic oscillators (as in the case of all cooperator state) may not lead to synchronization, occasional uncoupling between the oscillators (as in the case of a mixed cooperator-defector state) may induce synchronization~\cite{2008_HS_C,2009_CQH_PRE,2015_SMDCT_PRL,2016_ZZGLB_SR,2016_TSMTS_Chaos,2016_SCWNT_SR,2018_LSCW_PRE}. The mixed state is what we are more interested here anyway because we want to know whether global cooperation can emerge in our model starting from a mixed state and how the cooperation co-evolves with the synchronization. A straightforward useful definition, in this context, is the degree of cooperation in the population that is just the fraction of cooperators (agents with strategy $s^i=1$)---denoted by $C$ througout the paper---in the population. Clearly, $C\in[0,1]$.
\begin{figure}
	\includegraphics[scale=0.24]{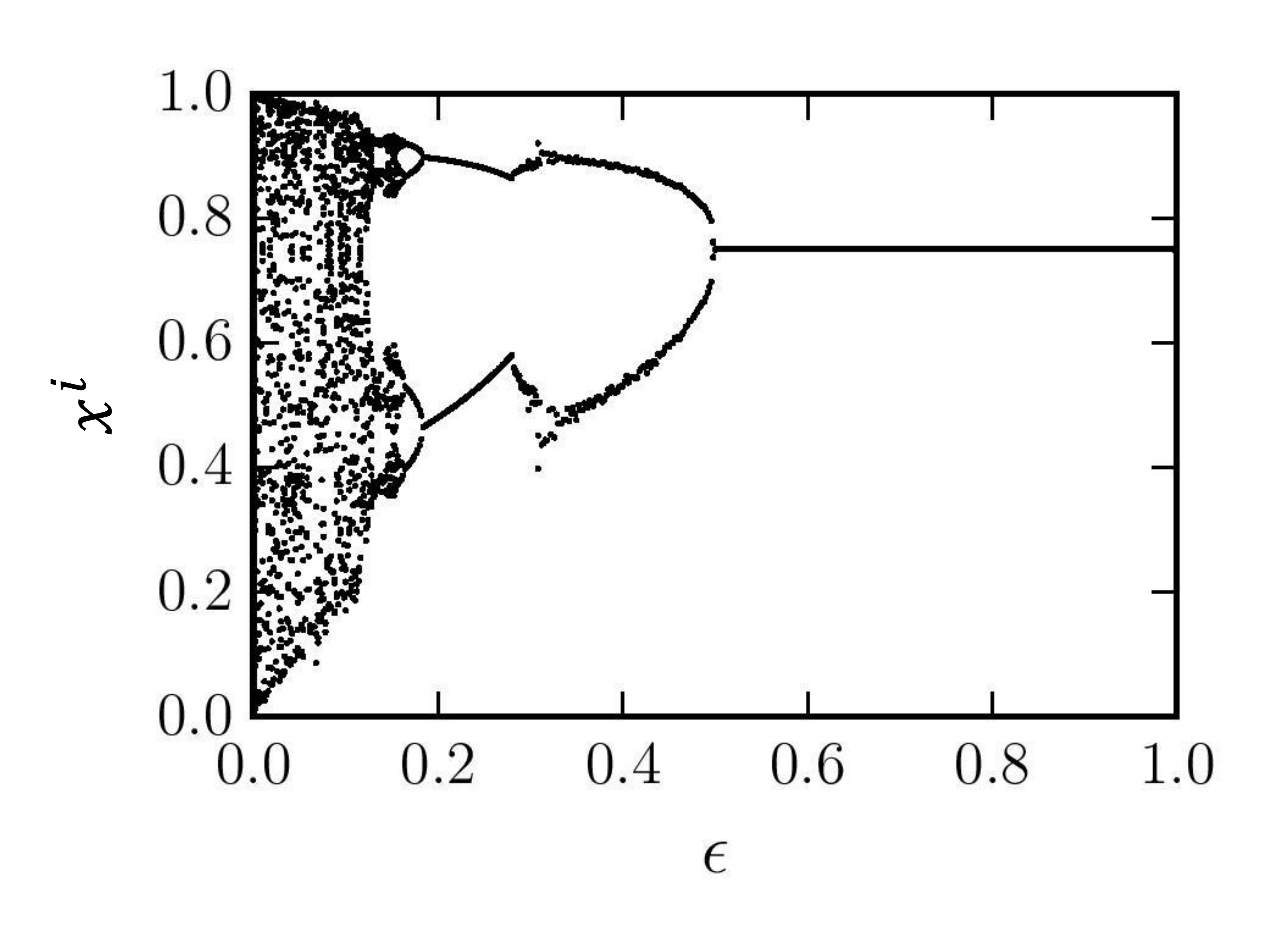}
	\caption{\textbf{Bifurcation diagram for a globally coupled network of cooperators.} We plot state variable ($x^i$) of a randomly chosen $i$th node in an all-to-all coupled network of 100 logistic maps after 2000 time steps for varying coupling strengths $\left(\epsilon\right)$. We find a fixed point for $\epsilon >0.5$. On turning down the coupling strength from 0.5 to 0, we find that period 2, period 4, and higher periods emerge, ultimately leading to chaos.}
	\label{bif_plt}
\end{figure}

Before proceeding further, we pause a bit to ponder upon the meaning of synchronization in the CML. The individual uncoupled agent's dynamics is chaotic. So, if the agents' dynamics synchronize once coupling is turned on, then a realizable synchronized state should be an attractor in the $N$-dimensional phase space, i.e., the phase space of the CML. The attractor could either be a homogeneous fixed point (i.e., $x^i_n=x^j_n$ $\forall i,j\in\{1,2,\cdots,N\}$ as $n\rightarrow\infty$) or possibly one of the non-fixed point attractors such as chaotic attractor and stable homogenous periodic orbit. For illustration, consider that all the agents are cooperators and there is all-to-all coupling in the CML---i.e., $\forall n,i,j$, $s^i_n=1$ and $a_{ij}=1-\delta_{ij}$ ($\delta_{ij}$ is the Kronecker delta) respectively---given by Eq.~(\ref{eq_rewired_strategy_network}).  The resulting equation has a trivial homogeneous fixed points, viz., $x^i=0$. However, it is not an attractor. The attractors depend on the value of the coupling constant. A bifurcation diagram corresponding this particular CML is depicted in Fig.~\ref{bif_plt}. We find that all the nodes are synchronized to a stable homogeneous fixed point $x^i=x^*=0.75$ when the coupling strength is greater than $0.5$. As we lower the coupling strength, the CML undergoes a period doubling route to chaos potentially followed by more chaotic regions and periodic windows. Thus, the CML's nodes can also synchronize onto the stable periodic orbits of various periods depending on the value of the coupling constant (e.g., period-$4$ at $\epsilon\approx0.17$) and, in principle, even onto a chaotic attractor. In all the cases, $r_G=1$ at all times (after sufficient transients are ignored) would indicate global synchronization but, of course, it cannot differentiate the different attractors.

\section{Numerics and Results}
\label{sec:R}
Our model allows for using any network topology for the CML. We exclusively use the WS model in which increasing a particular parameter---the rewiring probability, $p\in[0,1]$---one can transition from a regular network ($p=0$), to small-world networks, and finally to a random network ($p=1$). In WS model, one builds a network topology by starting with a undirected ring lattice with $N$ nodes, each having $k$ edges ($k/2$ on each side of a node). Subsequently, proceeding in a anticlockwise manner (say), the left nearest neighbour (along the ring) connections of each node are rewired with a probability $p$ to a random node while avoiding self and multiple connections. Once the nearest neighbours of all the nodes are exhausted, one again considers every node sequentially and randomly rewires the next nearest left neighbour (along the ring) links of each node with a probability $p$. This process is repeated until all the links of the ring lattice one started with have been considered for rewiring and finally we arrive at a network with $N$ nodes and average degree $k$. In this paper, neighbours of an agent mean all other agents who have a connecting edge with the agent.

We intend to study how the global synchronization parameter and the degree of cooperation behave in the CML, thus created, as we change its network topology by tuning the rewiring probability and the degree of the network. Interestingly, we find that our results pertaining to such studies are in sharp contrast with the similar studies~\cite{2018_Liu_Wu_Guan_EPL, 2018_Yang_Han-Xin_Zhou} where the uncoupled agent dynamics is of the Kuramoto oscillator and not a chaotic map.

In what follows, unless otherwise specified, we present our results using, the WS networks with 100 nodes and average degrees up to 98. We evolve the system for 2000 time steps that are enough to reach the steady state solutions. We do check that the final results are qualitatively robust against varying network size. All the reported values of the global synchronization parameter and the degree of cooperation are calculated using the data at the final time step. Also, they are averaged over 40 independent realizations achieved using 40 different random initial conditions and as many realizations of the network in Eq.~(\ref{eq_rewired_strategy_network}), and we use $\langle r_G\rangle$ and $\langle C\rangle$ to denote them respectively. Similarly, wherever appropriate, we use $\langle k\rangle$ for the average degree in the light of aforementioned averaging.
\subsection{Critical coupling strength for synchronization}
\label{sec:crit}
\begin{figure}[h]
	\includegraphics[scale=1]{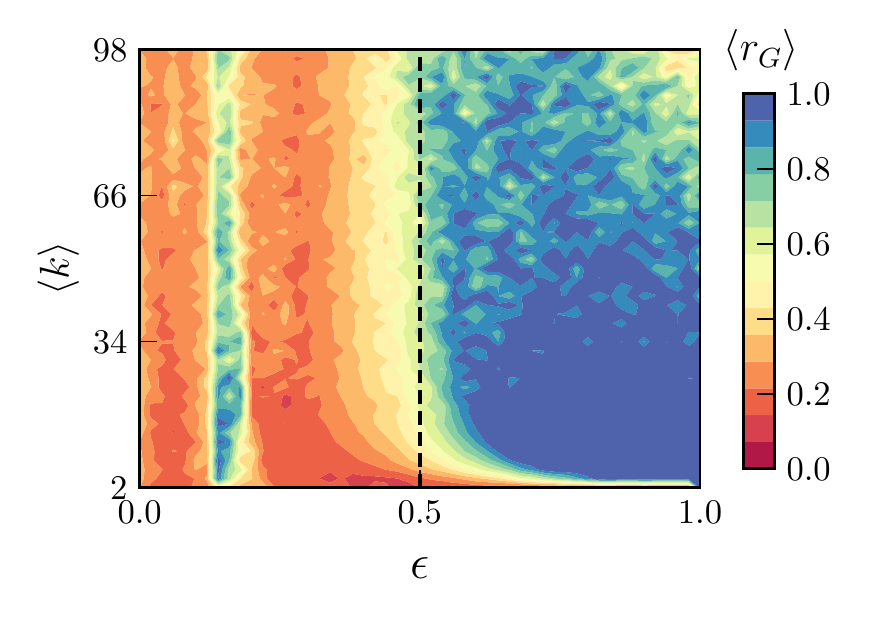}
	\caption{\textbf{Global order parameter with varying coupling strength and average degree.} We plot the average global order parameter $\left(\langle r_G\rangle\right)$ for a WS network (details in the text), as a function of the average degree $\left(\langle k\rangle\right)$ and the coupling strength $\left(\epsilon\right)$. We fix the relative cost $\left(\alpha\right)$ to 0.01 and rewiring probability ($p$) to 0.2.  The vertical dashed black line corresponds to the analytically estimated value of the critical coupling strength $\left(\epsilon_{\rm crit}\right)$ in the limit $N,\langle k\rangle\rightarrow\infty$. 
	}
	\label{fig_order_deg}
\end{figure}
The CML is quite an analytically challenging dynamical system. So the natural question, that what the minimum coupling strength should be so that synchronization is effected, is not easy to answer analytically. Nevertheless, there is a case where this question can be answered and the critical coupling strength, $\epsilon_{\rm crit}$, beyond which a stable synchonized state---a homogenous fixed point attractor---is realized (e.g., $\langle r_G\rangle=1$), apparently turns out to be a lower bound for the other cases studied in this paper. The case happens to be the limit of $N,\langle k\rangle\rightarrow\infty$ in the CML where all agents adopt the strategy of cooperation in the synchronized state.

Thus, let's recall Eq.~(\ref{eq_rewired_strategy_network}) with all $s^i=1$. Since we are looking for a nontrivial homogeneous fixed point, the only possibility is $x^i=x^*=0.75$ $\forall i$. We now must look for the minimum value of coupling constant (actually, $\epsilon_{\rm crit}$) such that the fixed point is stable. To this end, we perform standard linear stability analysis by putting $x^i_{n}=x^{*}+h_n^i$, $h_n^i$ being the infinitesimal perturbation, in Eq.~(\ref{eq_rewired_strategy_network}), and on expanding to the first order terms, we arrive at,
\begin{eqnarray}
	\label{1storder}
	h_{n+1}^i=&4\textcolor{black}{\left(1-2x^{*}\right)}\left(1-\epsilon\right)h_n^i+\frac{\epsilon}{\langle k\rangle}\sum_{j=1}^{N}a_{ij}h_n^j.
\end{eqnarray}
Note we have replaced $k_i$ with $\langle k\rangle$ because we are interested in an ensemble of networks with the same number of nodes and the same average degree. Subsequently, we use the Fourier expansion for the perturbation---$h_n^i=\sum_q \tilde{h}_n^{(q)} \exp\left(\sqrt{-1}iq\right)$, and substitute it in Eq.~(\ref{1storder}) to get for every $q$th Fourier mode,
\begin{eqnarray}
	\label{f_transform}
	\frac{\tilde{h}_{n+1}^{(q)}}{\tilde{h}_n^{(q)}}=&& 4\left(1-2x^*\right)\left(1-\epsilon\right).
\end{eqnarray}
Here, we have imposed the condition $\langle k\rangle\rightarrow\infty$. Evidently, the synchronized state is stable if $|{\tilde{h}_{n+1}^{(q)}}/{\tilde{h}_n^{(q)}}|<1$ or $|4(1-2x^*)(1-\epsilon)|<1$. Since $x^*=0.75$ and $0\le\epsilon\le1$, $\epsilon_{\rm crit}=0.5$, i.e., the CML spatiotemporally synchronizes for any value of coupling constant greater than half. 

It is worth remarking that since the estimation of $\epsilon_{\rm crit}$ has not explicitly required any mention of the type of network, this threshold value of the coupling strength should hold good for any network with  $N,\langle k\rangle\rightarrow\infty$.  Furthermore, although analytically quite challenging, in principle, such an estimation of $\epsilon_{\rm crit}$ may be possible for synchronization onto a homogeneous $n$-period orbit attractor, $(x^{1*},x^{2*},\cdots,x^{n*})$, such that all the nodes synchronously fluctuate between these $n$ values for each agent's state as time tends to infinity. In fact, numerically we do find (refer to Fig.~\ref{fig_order_deg}), the system synchronizes onto a 4-period orbit around $\epsilon\approx0.17$ that is interestingly also seen in Fig.~\ref{bif_plt}. The other synchronization regime---a synchronization onto the homogeneous stable fixed point---is found at much higher coupling strengths. We note that as $\langle k\rangle$ increases the analytical estimate of $\epsilon_{\rm crit}=0.5$ matches reasonably well with the numerical results. For all practical purpose, $\langle k\rangle\sim30$ is large enough for the estimate to be considered valid. It may also be observed that in Fig.~(\ref{fig_order_deg}), $\langle r_G\rangle$ is not exactly equal to unity beyond $\epsilon_{\rm crit}$ not only because $\langle k\rangle$ is not strictly infinity, but also because the averaging is over only a finite number of realizations.
\subsection{Effect of changing rationality}
\label{sec:r}
\begin{figure*}[htbp]
	\includegraphics[scale=0.9]{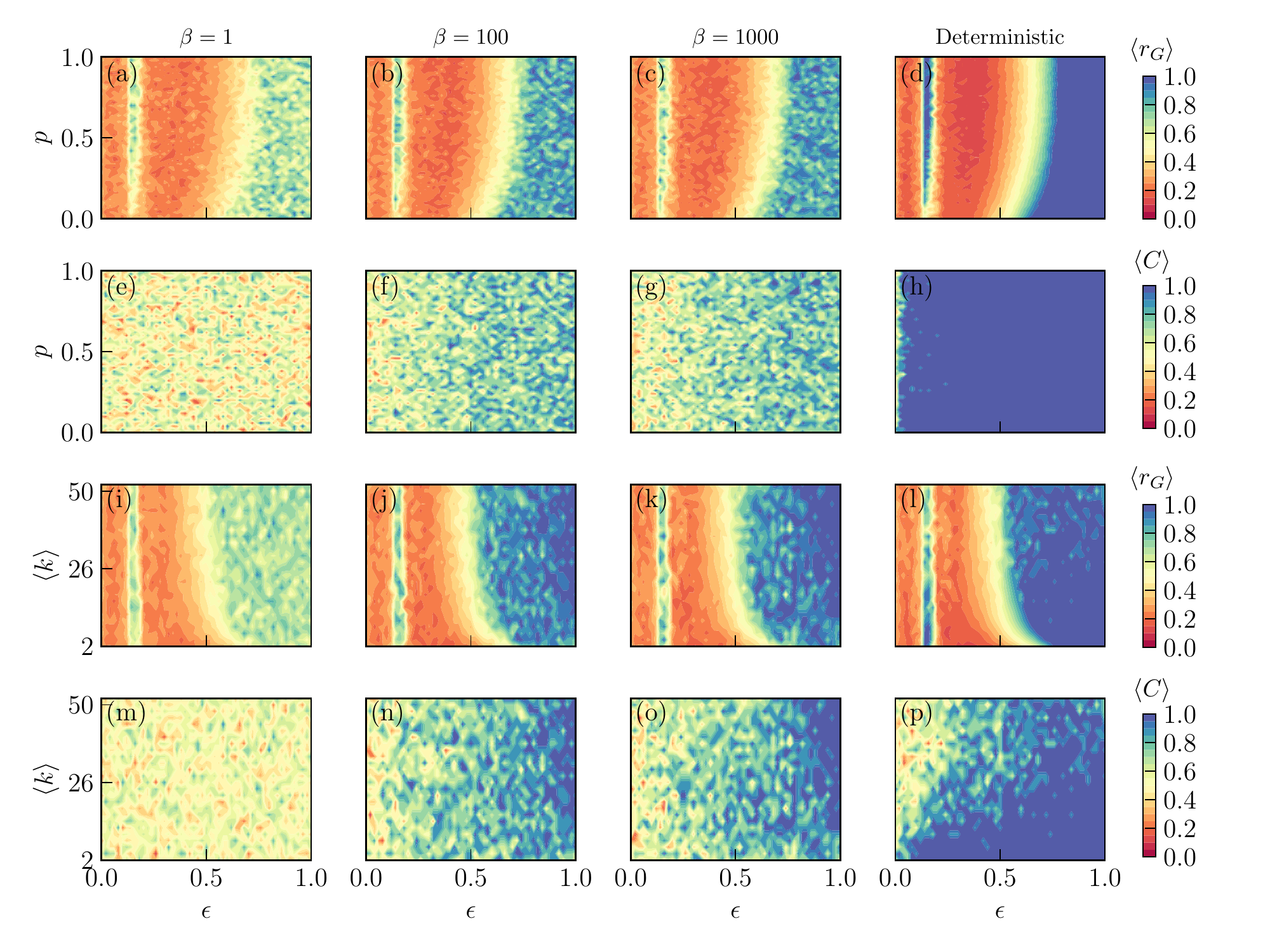}
	\caption{\textbf{Order parameter and cooperation for different strategy update rules.} We plot in (a)-(d) and (i)-(l) the average global order parameter $\langle r_G\rangle$ and in (e)-(h) and (m)-(p) the average fraction of cooperators $\langle C \rangle$ in the CML of the logistic maps. We use a WS network of 100 nodes and $r_G$ and $C$ are averaged over 40 independent realizations. The relative cost $\alpha$ is kept fixed at 0.01 for all the subplots. The right most subplot, viz., (d), (h), (l) and (p), in each row is obtained with the deterministic strategy exchange rule. The rest of three plots in every row is obtained using the Fermi strategy update rule having rationality factor $\left(\beta\right)= 1, 100, \textrm{ and} ~1000$ respectively, from left to right. Subplots (a)-(h) are obtained with the average degree ($\langle k\rangle$) fixed at $6$ while subplots (i)-(p) are obtained for a fixed rewiring probability, $p=0.5$.}
	\label{order_para}
\end{figure*}
Another important parameter on which the effectiveness of the critical coupling strength, estimated in the immediately preceding subsection, depends is the strategy update rule. 

It may be recalled that the stochastic Fermi strategy update rule, which we adopt in this paper, has a rationality factor $\beta$ that quantifies the degree of rationality of the agents. Its meaning becomes crystal clear when one notes that an infinitely rational ($\beta=\infty$) $i$th agent definitely chooses the strategy of one of the randomly chosen $j$th neighbouring agent if the $j$the agent obtained more payoff than the $i$th agent. On the other hand, the infinitely rational $i$th agent never chooses the strategy of one of the $j$th neighbouring agent if the $j$the agent has comparatively less payoff. Similarly, in case the $i$th agent is completely irrational ($\beta=0$), it chooses the strategy of the randomly chosen $j$th neighbouring agent with probability equal to one-half. Thus, $\beta$ is an apt parameter to characterize the degree of rationality, because with the increase in its value, the probability of choosing the strategy of the neighbouring agent with higher payoff increases monotonically. It is obvious that the stochastic update rule with $\beta=\infty$ is practically equivalent to the deterministic rule---the synchronous unconditional imitation---also adopted in this paper (see Sec.~\ref{sec:model})---only difference being that in the former an agent only compares its payoff with that of a randomly chosen neighbour, while in the latter an agent compares its payoff with all its neighbours.

In Fig.~\ref{order_para}, we depict $\langle r_G\rangle$ and $\langle C\rangle$ as a function of the rewiring probability and the coupling constant (the average degree fixed at $6$), and also as a function of the average degree and the coupling constant (the rewiring probability fixed at $0.5$) for both the the stochastic strategy update rule (with $\beta=1$, $100$, and $1000$) and the deterministic strategy update rule. The relative cost $\left(\alpha\right)$  is kept fixed at $0.01$. We see no qualitative difference in the results obtained from stochastic or deterministic strategy update rules as far as degree of synchronization's dependence on the rewiring probability, the coupling constant, and the average degree is concerned.  Another fact we note is that, in line with our discussion above, the plots for both $\langle r_G\rangle$ and $\langle C\rangle$ found using the stochastic strategy update rule tend towards the corresponding ones obtained using the deterministic strategy update rule as we increase the rationality factor. 

We further observe that, having kept all other parameters fixed , the degrees of cooperation and synchronization achieved using  the unconditional imitation rule are generally higher than in the Fermi strategy update rule. It is consistent with what is known in the literature~\cite{2009_RCS_PLR}. However, the average qualitative behaviour of the dynamics usually remains the same irrespective of which of the two update rules is adopted~\cite{2017_Antonioni_Cardillo_PRL}. Furthermore, the direct correspondence of the highly cooperative population and its being spatiotemporally synchronized is best captured in the deterministic strategy update rule because, as expected, less rational players do not play strategically but rather randomly choose a strategy leading to a mixed cooperator-defector state.  Thus, henceforth, we exclusively work with the the unconditional imitation rule that has the added benefit that being a non-stochastic scheme, it facilitates a sharper boundary between the completely synchonized ($\langle r_G\rangle=1$) regions and partially synchronized ($\langle r_G\rangle<1$) or unsynchronized ($\langle r_G\rangle=0$) regions. Similar relatively sharper boundary is presented by the unconditional imitation rule in the plots for cooperation ($\langle C\rangle$) as well. 
\subsection{ Effect of changing rewiring probability}
\label{sec:p}
\begin{figure}[h]
	\includegraphics[scale=1]{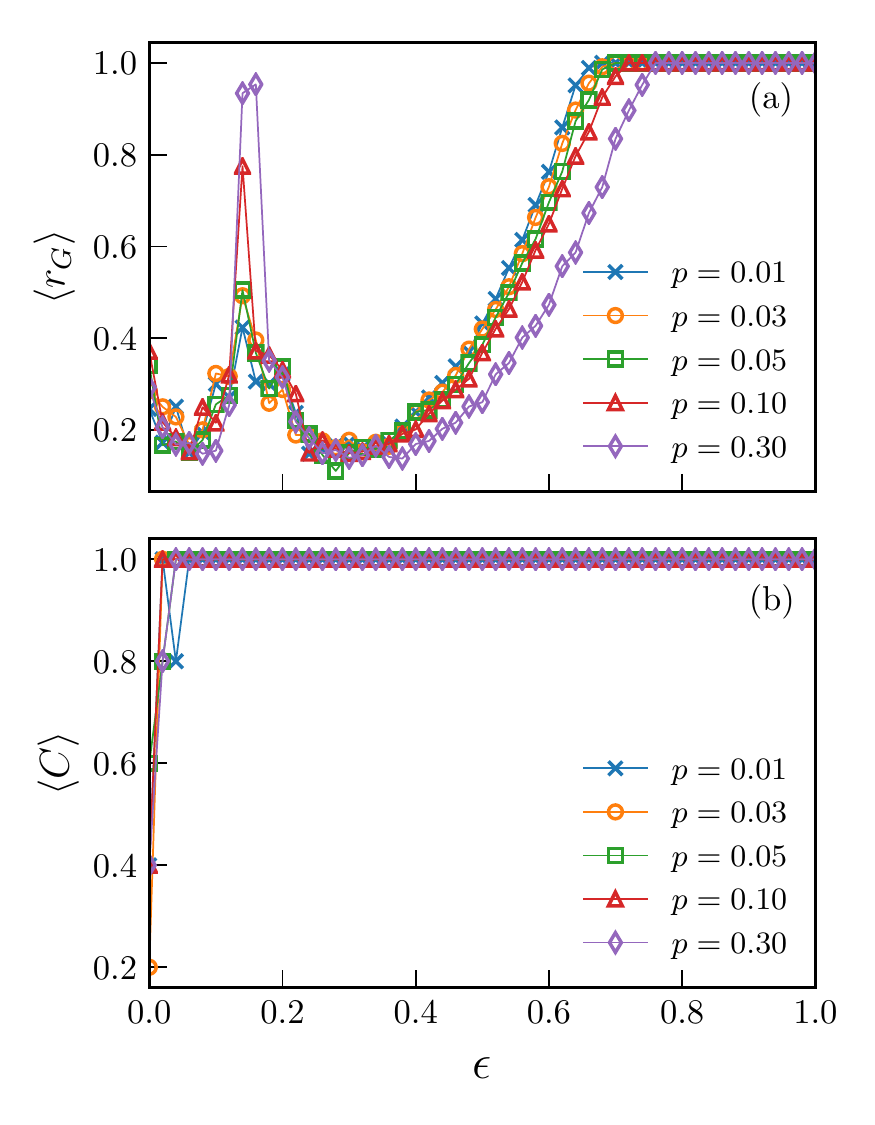}
	\caption{\textbf{Order parameter and cooperation for different rewiring probabilities.} We plot (a) average global order parameter $\left(\langle r_G\rangle \right)$ and (b) average cooperation $\left(\langle C\rangle\right)$ as a function of coupling strength, $\epsilon$. The average degree and relative cost are fixed at 6 and 0.01 respectively. The blue, the orange, the green, the red, and the violet lines correspond to the following values of the rewiring probability: $p= 0.00, 0.01, 0.05, 0.10,\textrm{and }0.30$ respectively.}
	\label{line_rewiring}
\end{figure}

In order to gain insight into the effect of the rewiring probability, $p$, on the coevolution of the cooperation and the synchronization, we present in Fig.~\ref{line_rewiring} how they change with the coupling strength for various representative rewiring probabilities. We observe that the CML reaches a relatively higher level of the global order parameter for a very narrow range of the coupling strength around $\epsilon\approx0.17$ where one finds that the system synchronizes onto a 4-period orbit. On further increasing the coupling strength, the system desynchronizes until respective values of $\epsilon_{\rm crit}$ are reached and beyond which complete synchronization onto the homogeneous fixed point takes place. Thus, {we see that the CML is completely synchronized---regardless of the rewiring probability---beyond some $\epsilon=\epsilon_{\rm crit}$ that increases with increasing $p$}. This result is at odds with the case of the Kuramoto model governed agent dynamics where the system attains complete synchronization only for relatively high rewiring probabilities and moreover the threshold coupling strength decreases with increasing rewiring probability~\cite{2018_Liu_Wu_Guan_EPL}. 

Except for the case where the coupling between the agents is very small (i.e., $\epsilon\rightarrow 0$), the CML (with any value of $p$) reaches full cooperation state whether or not it is synchronized. An understanding of the mechanism leading to this phenomenon reveals the nature of the local dynamics in the CML. First, consider the case when the coupling is very weak ($\epsilon\rightarrow 0$). In this case, the agents may be thought to be evolving chaotically and independently, and there is no hope of synchronization among them. Consequently, any agent is not synchronized with its neighbours, let alone with the rest of the non-neighbouring agents. This immediately implies that the benefit (see Eq.~\ref{eq:benefit}) is very small and on average same for all the agents---whether cooperators or defectors. The cost (see Eq.~\ref{eq:cost}; $\alpha=0.01$) is also small and but on average same only for the cooperators; the defectors, by construction, incur zero cost. Hence, the payoff of the cooperators is smaller than that of the defectors leading to the obvious scenario that under the unconditional imitation rule, all the agents would adopt the defection strategy sooner or later.

Next consider the case, $\epsilon>\epsilon_{\rm crit}$, where the parameters of the CML is so arranged that it is driven towards a completely synchronized state which is a homogeneous fixed point. In the basin of attraction of such a globally synchronized state, consider a close-by neighbouring state. In that neighbouring state, the benefit obtained by any cooperating agent is high (order $1$) since, by construction, the benefit is a measure of synchronization between the agent and its neighbours. The the cost should be low as is evident from its definition and the fact that all $x^i$s are almost same. The payoff of a cooperator, hence, should be high compared to that of a defector who is dynamically uncoupled from the neighbours. Naturally, under the unconditional imitation rule, all the agents would adopt the cooperator strategy as the CML's state rushes towards the stable synchronization state. Analogous argument goes for the region of synchronization onto stable homogenous period-4 orbit.

Lastly, consider the case: $0\ll\epsilon<\epsilon_{\rm crit}$ (ignoring the region of synchronization onto period-4 orbit). While the CML is not synchronized, the degree of cooperation in it is unity. This can be attributed to the following mechanism: Although the CML is not completely synchronized, the non-zero value of $\langle r_G\rangle$ suggests that it is possible for a few agents whose local synchronization parameter (which actually is the benefit)---which measures their dynamics' closeness with their respective neighbours'---to be much higher than its global counterpart. So, within such groups, the unconditional imitation rule facilitates the agents to adopt the cooperation strategy eventually. Being locally synchronized, such cooperators in the groups have higher payoff and hence, any defector---who has a high probability of being connected with one of the so abundant cooperators---must also switch to the cooperation strategy because of the unconditional imitation rule.

\subsection{Effect of changing degree}
\label{sec:k}

\begin{figure*}[htbp]
	\includegraphics[scale=0.8]{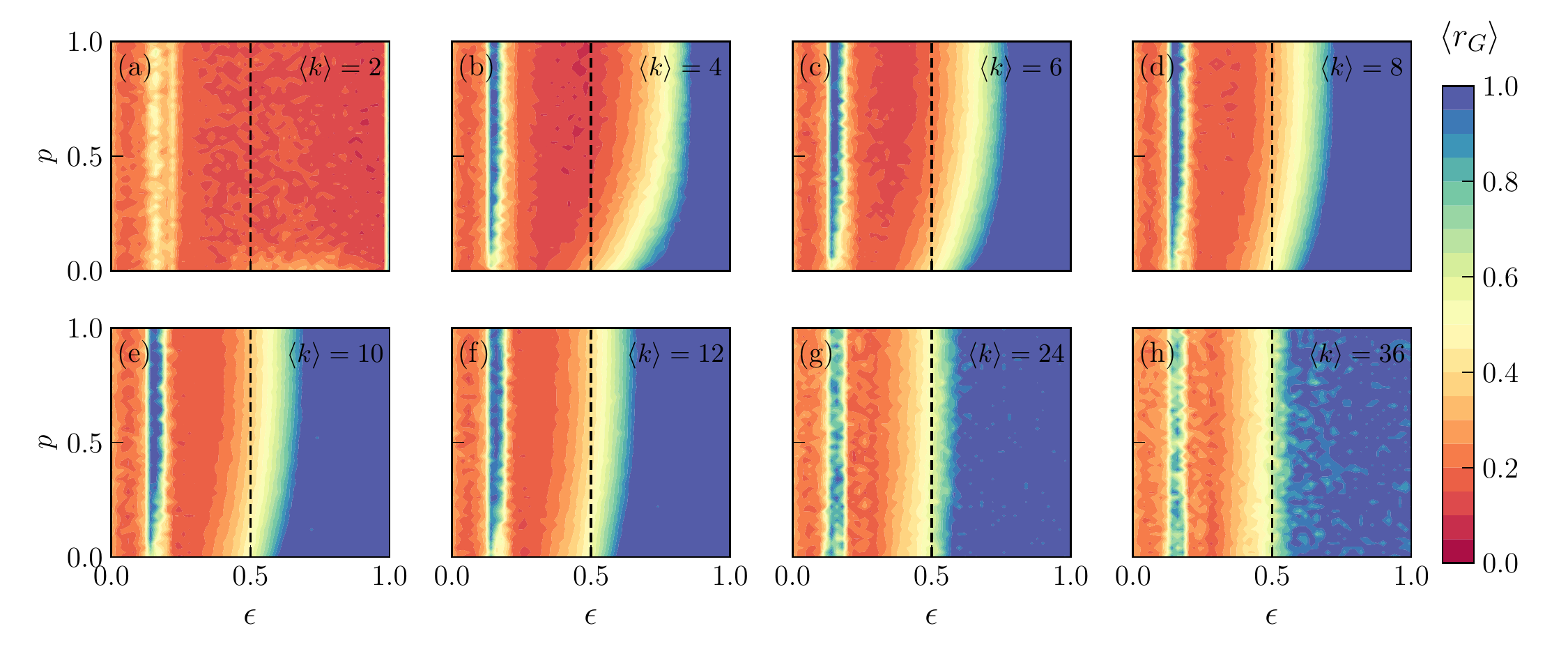}
	\caption{\textbf{Order parameter for varying rewiring probability and degree.} We plot the average order parameter ($\langle r_G\rangle$), for a CML of 100 nodes averaged over 40 independent realizations, as a function of the coupling strength ($\epsilon$) and the rewiring probability ($p$). The relative cost $\alpha$ is fixed at 0.01. Subplots (a)-(h) are obtained with average degrees ($\langle k\rangle$) 2, 4, 6, 8, 10, 12, 24, and 36, respectively. The vertical dashed black lines correspond to the analytically estimated value of the critical coupling strength, $\epsilon_{\rm crit}$, in the limit $N,\langle k\rangle\rightarrow\infty$.}
	\label{fig_coop_fixed}
\end{figure*}
Continuing from the last subsection, we note in Fig.~\ref{order_para}(p) that, for a given rewiring probability, the cooperation level decreases in the non-synchronized states of the CML as the average degree increases. Evidently, in a mixed cooperator-defector state of the population, since a defector incurs no cost, a cooperator would switch over to the defector strategy as per the rules of strategy update in case one such defector happens to be in the neighbourhood of the cooperator. The chance of such a defector being in the neighbourhood should be more if the degree of the node having cooperator is higher. Hence, the networks with higher average degree would decrease the cooperation. 

The effect on synchronization is very interesting. In contrary to the case of non-chaotic agent dynamics where with a small increase in the degree of the network, the population becomes completely unsynchronized~\cite{2018_Yang_Han-Xin_Zhou}, the CML under consideration has a more nontrivial behaviour. Numerics depict in Fig.~\ref{fig_coop_fixed} that as the rewiring probability increases, $\epsilon_{\rm crit}$ also increases for relatively small average degree. As we increase the average degree, we find $\epsilon_{\rm crit}$ becomes independent of the rewiring probability and converges towards $\epsilon=\epsilon_{\rm crit}=0.5$ as it should. Note that as the degree increase across the subplots in the figure (see also Fig.~\ref{fig_order_deg}), the intensity of the colours towards diminishes validating an increase in the level of the synchronization with $\langle k\rangle$ for $\epsilon<0.5$. However, for $\epsilon>0.5$, on average, there is a decrease in the level of the synchronization with the average degree.

The rise in overall synchronization with an increase in degree~\cite{2004_LGH_PRE}, for the case $\epsilon<0.5$, can be attributed to the increase in the level of synchronization among local clusters of cooperating agents. Of course, the coupling strength in this region is not strong enough to bring about global synchronization. As far as the region $\epsilon>0.5$ is concerned, the coupling strength is strong enough to bring about global synchronization (see Fig.~\ref{fig_coop_fixed}). As mentioned in the beginning of this subsection, due to the noncooperative behaviour encouraged by the internodal game, there is a decrease in the number of cooperators with an increase in the degree of the network. The direct consequence of this loss of cooperators is the loss in complete synchronization as more agents choose not to interact with their neighbours. Although the number of cooperators is reduced, it does not go down to zero. This implies that there are still local clusters of cooperators. These local clusters of cooperators, coupled with a high value of coupling strength, gives a high value of the global order parameter in $\epsilon>0.5$ region as compared to $\epsilon<0.5$ region for the same degree. Intriguingly, the internodal game---though successful in breaking the complete synchronization of the system for $\epsilon>0.5$---is unable to break the small clusters of the cooperating agents that exist throughout the region of $0<\epsilon<1$. These clusters seem to be held together by the degree of the network and the dynamics of the agents placed at the node. The degree of synchronization never diminishes to zero entirely in our model and, thus, the type of agent dynamics appears to be crucial for the sustenance of the synchronized state having prevalence of the cooperators.
\subsection{Effect of introducing delay}
\label{sec:d}
Before we discuss and conclude the results obtained in this paper, we turn on to another aspect of the model: What if there is a delay in  updating in strategy by the agents? A delay in updating strategy has the direct effect that the strategic game in the CML has relatively more intermittent effect on the dynamics of the agents' individual and collective states. It means that in a mixed cooperator-defector state, a cooperator resist from adopting the defector strategy for a longer time, and hence the coupling between the agent and its neighbours remain turned on for longer time facilitating synchronization. Thus, all other parameters kept fixed, larger delay should mean synchronization---and emergence of cooperation with it---at the smaller values of the coupling strength.

\begin{figure}[htbp]
	\includegraphics[scale=0.9]{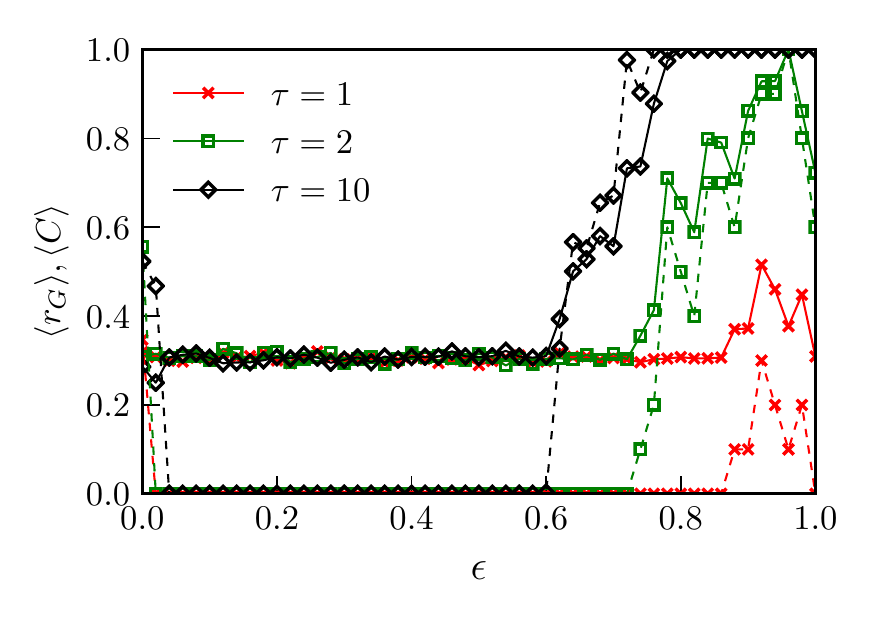}
	\caption{\textbf{The effect of delay in strategy update on synchronization and cooperation.} We plot the order parameter, $\langle r_G\rangle$ (solid lines) and the associated average cooperation, $\langle C\rangle$ (dashed lines) as a function of the coupling strength for different delay times---$\tau= 1\textrm{ (red)}, 2\textrm{ (green)},\textrm{ and }10\textrm{ (black)}$. The fixed relative cost, the average degree, and the rewiring probability are taken to be 1, 6, and 0.5, respectively.}
	\label{delay_plot}
\end{figure}

We implement the delay in the unconditional imitation strategy update rule in our system by allowing for a strategy update at each time step with a probability of $\frac{1}{\tau}$ ($\tau\in[1,\infty)$), where $\tau=1$ corresponds to the usual case of the strategy update at every time step; the higher the value of $\tau$, the more delayed the strategy update is. Fig.~\ref{delay_plot} illustrates that when we incorporate a delay in our strategy update, there is a simultaneous rise in the degrees of synchronization and cooperation at a relatively lower coupling strength. Moreover,  the coordinated falls in $\langle r_G\rangle$ and $\langle C \rangle$, that occur at a higher value of the coupling strength, are completely halted is the delay is sufficiently high. This result is at odds with the case of non-chaotic nodal dynamics~\cite{2018_Liu_Wu_Guan_EPL}, where the system's degrees of cooperation and synchronization begin to rise from a very low initial values at the same coupling strength regardless of the delay time; but the configuration with a higher delay time, loses synchronization at a relatively higher value of the coupling strength.

Despite the differences in the details between the two systems---one with chaotic agent dynamics and the other with non-chaotic dynamics---we conclude that, in general, delay aids in the co-emergence of synchronization and cooperation for both chaotic and non-chaotic agent dynamics.

In passing we remark that as long as $\alpha$ is very low, introducing delay is not expected to have much effect on the degrees of synchronization and cooperation as compared to the case when there is no delay in the strategy update. In Fig.~\ref{line_rewiring}, we can observe that even at a very low values of $\epsilon$, at low cost value ($\alpha=0.01$), the system has complete cooperation. Thus, introducing delay cannot increase the fraction of cooperators any further. However, high $\alpha$ (e.g., $\alpha=1$ in Fig.~\ref{delay_plot}) implies a high cost of cooperation and we may expect the system to desynchronize in the no-delay case ($\tau=1$) where the system is otherwise synchronized for low $\alpha$. It is important to note that regardless of the high value of $\alpha$, the cost can be brought close to zero if the agents do not change their states much with each time step. This happens when the agents are (almost) synchronized, i.e., when the coupling strength is high enough. Therefore, if the cooperators are made to resist from adopting the defector strategy by introducing delay, then we can see the system to synchronize again at high values of coupling strength despite the high value of $\alpha$ as seen in Fig.~\ref{delay_plot}. We also note in Fig.~\ref{delay_plot} (cf.~Fig~\ref{line_rewiring}) that the region of high degree of synchronization onto 4-period orbit around $\epsilon\approx 0.17$ has disappeared for all values of delay. This is because $\alpha$ is high and the low value of $\epsilon$ in this region cannot bring the agents' states close together quickly enough before the strategy update occurs.
\section{Discussions and Conclusions}
\label{sec:DnC}

In this paper, we have investigated the effect of chaotic agent dynamics on the coevolution of cooperation and synchronization in the networks formed using the WS algorithm. Specifically, we have placed logistic maps at each node of the network and have formed a CML that can show period doubling route to chaos with the decrease in the coupling strength. Subsequently, we have studied the effects of the strategy update rule, the rationality of the agents, the coupling strength, the rewiring probability of the network, and the average degree of the network on the coevolution of cooperation and synchronization. We have also analytically estimated---and numerically validated---a lower bound of the critical coupling constant beyond which global synchronization on to a common fixed state for all the agents may be affected. An aspect that is unique to our model is the possibility of more exotic synchronized states like chaotic state and periodic state. An example for the latter---a $4$-period orbit---has been highlighted in the paper. We have shown that the CML is completely synchronized, irrespective of what the rewiring probability is, beyond critical coupling strength that increases with the value of rewiring probability. When we increase the degree, we observe that the population of the agents have high values of the global coupling parameter even for very high average degree of the network. Furthermore, any time delay in the implementation of the strategy update rule induces enhancement of the coevolution of cooperation and synchronization at comparatively lower coupling strengths. These results are quite different from what is known in the literature~\cite{2017_Antonioni_Cardillo_PRL,2018_Liu_Wu_Guan_EPL, 2018_Yang_Han-Xin_Zhou} of the coevolution of cooperation and synchronization with non-chaotic agent dynamics studied by placing the Kuramoto oscillators at each node of the network. Thus, we have successfully highlighted in this paper that the interesting phenomenon of the coevolution of cooperation and synchronization is crucially dependent on whether the uncoupled agent dynamics is chaotic or not.

This paper initiates a set of potentially insightful investigations into the further implications of different chaotic dynamics of the agents on the emergence of cooperation in a population. One such avenue is when the uncoupled dynamics at each node of the resulting CML can be modelled using the replicator map~\cite{2018_PMS_C,2020_MC_JTB} that may even be chaotic depending of the payoff matrix elements in it. Each node can either be treated as an individual or a group of individuals (deme); in the latter case, the dynamics of the CML would model the intergroup dynamics in a population. Thus, it presents an exciting opportunity of studying~\cite{2007_MB} group-selection models and the intrademic cooperation within the paradigm of the coevolution of (interdemic) cooperation and synchronization.

\section*{ Acknowledgments} 
The authors are thankful to Archan Mukhopadhyay and Samriddhi Shankar Ray for helpful discussions. 
\section*{AIP Publishing data sharing policy} 
\textcolor{black}{The data that support the findings of this study are available from the corresponding author upon reasonable request.}
%
\bibliography{Chattopadhyay_etal_manuscript.bib}

\begin{thebibliography}{65}%
\makeatletter
\providecommand \@ifxundefined [1]{%
 \@ifx{#1\undefined}
}%
\providecommand \@ifnum [1]{%
 \ifnum #1\expandafter \@firstoftwo
 \else \expandafter \@secondoftwo
 \fi
}%
\providecommand \@ifx [1]{%
 \ifx #1\expandafter \@firstoftwo
 \else \expandafter \@secondoftwo
 \fi
}%
\providecommand \natexlab [1]{#1}%
\providecommand \enquote  [1]{``#1''}%
\providecommand \bibnamefont  [1]{#1}%
\providecommand \bibfnamefont [1]{#1}%
\providecommand \citenamefont [1]{#1}%
\providecommand \href@noop [0]{\@secondoftwo}%
\providecommand \href [0]{\begingroup \@sanitize@url \@href}%
\providecommand \@href[1]{\@@startlink{#1}\@@href}%
\providecommand \@@href[1]{\endgroup#1\@@endlink}%
\providecommand \@sanitize@url [0]{\catcode `\\12\catcode `\$12\catcode
  `\&12\catcode `\#12\catcode `\^12\catcode `\_12\catcode `\%12\relax}%
\providecommand \@@startlink[1]{}%
\providecommand \@@endlink[0]{}%
\providecommand \url  [0]{\begingroup\@sanitize@url \@url }%
\providecommand \@url [1]{\endgroup\@href {#1}{\urlprefix }}%
\providecommand \urlprefix  [0]{URL }%
\providecommand \Eprint [0]{\href }%
\providecommand \doibase [0]{http://dx.doi.org/}%
\providecommand \selectlanguage [0]{\@gobble}%
\providecommand \bibinfo  [0]{\@secondoftwo}%
\providecommand \bibfield  [0]{\@secondoftwo}%
\providecommand \translation [1]{[#1]}%
\providecommand \BibitemOpen [0]{}%
\providecommand \bibitemStop [0]{}%
\providecommand \bibitemNoStop [0]{.\EOS\space}%
\providecommand \EOS [0]{\spacefactor3000\relax}%
\providecommand \BibitemShut  [1]{\csname bibitem#1\endcsname}%
\let\auto@bib@innerbib\@empty
\bibitem [{\citenamefont {Smith}(1998)}]{1998_Smith}%
  \BibitemOpen
  \bibfield  {author} {\bibinfo {author} {\bibfnamefont {J.~M.}\ \bibnamefont
  {Smith}},\ }\href {https://www.xarg.org/ref/a/019850294X/} {\emph {\bibinfo
  {title} {The Major Transitions in Evolution}}}\ (\bibinfo  {publisher}
  {Oxford University Press, Oxford, UK},\ \bibinfo {year} {1998})\BibitemShut
  {NoStop}%
\bibitem [{\citenamefont {Nowak}\ and\ \citenamefont
  {Coakley}(2013)}]{2013_NC}%
  \BibitemOpen
  \bibinfo {editor} {\bibfnamefont {M.~A.}\ \bibnamefont {Nowak}}\ and\
  \bibinfo {editor} {\bibfnamefont {S.}~\bibnamefont {Coakley}},\ eds.,\ \href
  {\doibase 10.2307/j.ctvjnrscp} {\emph {\bibinfo {title} {Evolution, Games,
  and God: The Principle of Cooperation}}}\ (\bibinfo  {publisher} {Harvard
  University Press, Cambridge, MA},\ \bibinfo {year} {2013})\BibitemShut
  {NoStop}%
\bibitem [{\citenamefont {Bourke}(2011)}]{2011_Bourke}%
  \BibitemOpen
  \bibfield  {author} {\bibinfo {author} {\bibfnamefont {A.~F.}\ \bibnamefont
  {Bourke}},\ }\href {\doibase 10.1093/acprof:oso/9780199231157.001.0001}
  {\emph {\bibinfo {title} {Principles of Social Evolution}}}\ (\bibinfo
  {publisher} {Oxford University Press, Oxford, UK},\ \bibinfo {year}
  {2011})\BibitemShut {NoStop}%
\bibitem [{\citenamefont {Smith}(1982)}]{1982_Smith}%
  \BibitemOpen
  \bibfield  {author} {\bibinfo {author} {\bibfnamefont {J.~M.}\ \bibnamefont
  {Smith}},\ }\href {https://www.xarg.org/ref/a/0521288843/} {\emph {\bibinfo
  {title} {Evolution and the Theory of Games}}}\ (\bibinfo  {publisher}
  {Cambridge University Press, Cambridge, UK},\ \bibinfo {year}
  {1982})\BibitemShut {NoStop}%
\bibitem [{\citenamefont {Axelrod}(2006)}]{2006_Axelrod}%
  \BibitemOpen
  \bibfield  {author} {\bibinfo {author} {\bibfnamefont {R.}~\bibnamefont
  {Axelrod}},\ }\href {https://www.xarg.org/ref/a/0465005640/} {\emph {\bibinfo
  {title} {The Evolution of Cooperation: Revised Edition}}}\ (\bibinfo
  {publisher} {Basic Books, New York},\ \bibinfo {year} {2006})\BibitemShut
  {NoStop}%
\bibitem [{\citenamefont {Nowak}(2006)}]{2006_Nowak}%
  \BibitemOpen
  \bibfield  {author} {\bibinfo {author} {\bibfnamefont {M.~A.}\ \bibnamefont
  {Nowak}},\ }\bibfield  {title} {\enquote {\bibinfo {title} {Five rules for
  the evolution of cooperation},}\ }\href {\doibase 10.1126/science.1133755}
  {\bibfield  {journal} {\bibinfo  {journal} {Science}\ }\textbf {\bibinfo
  {volume} {314}},\ \bibinfo {pages} {1560--1563} (\bibinfo {year}
  {2006})}\BibitemShut {NoStop}%
\bibitem [{\citenamefont {Gintis}(2009)}]{2009_Gintis}%
  \BibitemOpen
  \bibfield  {author} {\bibinfo {author} {\bibfnamefont {H.}~\bibnamefont
  {Gintis}},\ }\href {https://www.xarg.org/ref/a/0691140510/} {\emph {\bibinfo
  {title} {Game Theory Evolving: A Problem-Centered Introduction to Modeling
  Strategic Interaction - Second Edition}}}\ (\bibinfo  {publisher} {Princeton
  University Press, Princeton, NJ},\ \bibinfo {year} {2009})\BibitemShut
  {NoStop}%
\bibitem [{\citenamefont {Perc}\ \emph {et~al.}(2017)\citenamefont {Perc},
  \citenamefont {Jordan}, \citenamefont {Rand}, \citenamefont {Wang},
  \citenamefont {Boccaletti},\ and\ \citenamefont {Szolnoki}}]{2017_PJRWS_PR}%
  \BibitemOpen
  \bibfield  {author} {\bibinfo {author} {\bibfnamefont {M.}~\bibnamefont
  {Perc}}, \bibinfo {author} {\bibfnamefont {J.~J.}\ \bibnamefont {Jordan}},
  \bibinfo {author} {\bibfnamefont {D.~G.}\ \bibnamefont {Rand}}, \bibinfo
  {author} {\bibfnamefont {Z.}~\bibnamefont {Wang}}, \bibinfo {author}
  {\bibfnamefont {S.}~\bibnamefont {Boccaletti}}, \ and\ \bibinfo {author}
  {\bibfnamefont {A.}~\bibnamefont {Szolnoki}},\ }\bibfield  {title} {\enquote
  {\bibinfo {title} {Statistical physics of human cooperation},}\ }\href
  {\doibase 10.1016/j.physrep.2017.05.004} {\bibfield  {journal} {\bibinfo
  {journal} {Phys. Rep.}\ }\textbf {\bibinfo {volume} {687}},\ \bibinfo {pages}
  {1 -- 51} (\bibinfo {year} {2017})}\BibitemShut {NoStop}%
\bibitem [{\citenamefont {Rapoport}\ and\ \citenamefont
  {Chammah}(1965)}]{1965_RC}%
  \BibitemOpen
  \bibfield  {author} {\bibinfo {author} {\bibfnamefont {A.}~\bibnamefont
  {Rapoport}}\ and\ \bibinfo {author} {\bibfnamefont {A.}~\bibnamefont
  {Chammah}},\ }\href {\doibase 10.3998/mpub.20269} {\emph {\bibinfo {title}
  {Prisoner's Dilemma}}}\ (\bibinfo  {publisher} {University of Michigan Press,
  Ann Arbor, MI.},\ \bibinfo {year} {1965})\BibitemShut {NoStop}%
\bibitem [{\citenamefont {Nowak}\ and\ \citenamefont {May}(1992)}]{1992_Nowak}%
  \BibitemOpen
  \bibfield  {author} {\bibinfo {author} {\bibfnamefont {M.~A.}\ \bibnamefont
  {Nowak}}\ and\ \bibinfo {author} {\bibfnamefont {R.~M.}\ \bibnamefont
  {May}},\ }\bibfield  {title} {\enquote {\bibinfo {title} {Evolutionary games
  and spatial chaos},}\ }\href {\doibase 10.1038/359826a0} {\bibfield
  {journal} {\bibinfo  {journal} {Nature}\ }\textbf {\bibinfo {volume} {359}},\
  \bibinfo {pages} {826--829} (\bibinfo {year} {1992})}\BibitemShut {NoStop}%
\bibitem [{\citenamefont {Ohtsuki}\ \emph {et~al.}(2006)\citenamefont
  {Ohtsuki}, \citenamefont {Hauert}, \citenamefont {Lieberman},\ and\
  \citenamefont {Nowak}}]{2006_OHLN_Nature}%
  \BibitemOpen
  \bibfield  {author} {\bibinfo {author} {\bibfnamefont {H.}~\bibnamefont
  {Ohtsuki}}, \bibinfo {author} {\bibfnamefont {C.}~\bibnamefont {Hauert}},
  \bibinfo {author} {\bibfnamefont {E.}~\bibnamefont {Lieberman}}, \ and\
  \bibinfo {author} {\bibfnamefont {M.~A.}\ \bibnamefont {Nowak}},\ }\bibfield
  {title} {\enquote {\bibinfo {title} {A simple rule for the evolution of
  cooperation on graphs and social networks},}\ }\href {\doibase
  10.1038/nature04605} {\bibfield  {journal} {\bibinfo  {journal} {Nature}\
  }\textbf {\bibinfo {volume} {441}},\ \bibinfo {pages} {502--505} (\bibinfo
  {year} {2006})}\BibitemShut {NoStop}%
\bibitem [{\citenamefont {Perc}\ and\ \citenamefont
  {Szolnoki}(2010)}]{2010_PS_B}%
  \BibitemOpen
  \bibfield  {author} {\bibinfo {author} {\bibfnamefont {M.}~\bibnamefont
  {Perc}}\ and\ \bibinfo {author} {\bibfnamefont {A.}~\bibnamefont
  {Szolnoki}},\ }\bibfield  {title} {\enquote {\bibinfo {title} {Coevolutionary
  games—a mini review},}\ }\href {\doibase
  https://doi.org/10.1016/j.biosystems.2009.10.003} {\bibfield  {journal}
  {\bibinfo  {journal} {BioSystems}\ }\textbf {\bibinfo {volume} {99}},\
  \bibinfo {pages} {109 -- 125} (\bibinfo {year} {2010})}\BibitemShut {NoStop}%
\bibitem [{\citenamefont {West}, \citenamefont {Mouden},\ and\ \citenamefont
  {Gardner}(2011)}]{2011_WCG}%
  \BibitemOpen
  \bibfield  {author} {\bibinfo {author} {\bibfnamefont {S.~A.}\ \bibnamefont
  {West}}, \bibinfo {author} {\bibfnamefont {C.~E.}\ \bibnamefont {Mouden}}, \
  and\ \bibinfo {author} {\bibfnamefont {A.}~\bibnamefont {Gardner}},\
  }\bibfield  {title} {\enquote {\bibinfo {title} {Sixteen common
  misconceptions about the evolution of cooperation in humans},}\ }\href
  {\doibase https://doi.org/10.1016/j.evolhumbehav.2010.08.001} {\bibfield
  {journal} {\bibinfo  {journal} {Evol. Hum. Behav.}\ }\textbf {\bibinfo
  {volume} {32}},\ \bibinfo {pages} {231 -- 262} (\bibinfo {year}
  {2011})}\BibitemShut {NoStop}%
\bibitem [{\citenamefont {Roca}\ and\ \citenamefont
  {Helbing}(2011)}]{2011_RH_PNAS}%
  \BibitemOpen
  \bibfield  {author} {\bibinfo {author} {\bibfnamefont {C.~P.}\ \bibnamefont
  {Roca}}\ and\ \bibinfo {author} {\bibfnamefont {D.}~\bibnamefont {Helbing}},\
  }\bibfield  {title} {\enquote {\bibinfo {title} {Emergence of social cohesion
  in a model society of greedy, mobile individuals},}\ }\href {\doibase
  10.1073/pnas.1101044108} {\bibfield  {journal} {\bibinfo  {journal} {Proc.
  Natl. Acad. Sci. U.S.A}\ }\textbf {\bibinfo {volume} {108}},\ \bibinfo
  {pages} {11370--11374} (\bibinfo {year} {2011})}\BibitemShut {NoStop}%
\bibitem [{\citenamefont {Hilbe}\ \emph {et~al.}(2018)\citenamefont {Hilbe},
  \citenamefont {{\v{S}}imsa}, \citenamefont {Chatterjee},\ and\ \citenamefont
  {Nowak}}]{2018_Hilbe}%
  \BibitemOpen
  \bibfield  {author} {\bibinfo {author} {\bibfnamefont {C.}~\bibnamefont
  {Hilbe}}, \bibinfo {author} {\bibfnamefont {{\v{S}}.}~\bibnamefont
  {{\v{S}}imsa}}, \bibinfo {author} {\bibfnamefont {K.}~\bibnamefont
  {Chatterjee}}, \ and\ \bibinfo {author} {\bibfnamefont {M.~A.}\ \bibnamefont
  {Nowak}},\ }\bibfield  {title} {\enquote {\bibinfo {title} {Evolution of
  cooperation in stochastic games},}\ }\href {\doibase
  10.1038/s41586-018-0277-x} {\bibfield  {journal} {\bibinfo  {journal}
  {Nature}\ }\textbf {\bibinfo {volume} {559}},\ \bibinfo {pages} {246--249}
  (\bibinfo {year} {2018})}\BibitemShut {NoStop}%
\bibitem [{\citenamefont {Mittal}, \citenamefont {Mukhopadhyay},\ and\
  \citenamefont {Chakraborty}(2020)}]{2020_MMC_PRE}%
  \BibitemOpen
  \bibfield  {author} {\bibinfo {author} {\bibfnamefont {S.}~\bibnamefont
  {Mittal}}, \bibinfo {author} {\bibfnamefont {A.}~\bibnamefont
  {Mukhopadhyay}}, \ and\ \bibinfo {author} {\bibfnamefont {S.}~\bibnamefont
  {Chakraborty}},\ }\bibfield  {title} {\enquote {\bibinfo {title}
  {Evolutionary dynamics of the delayed replicator-mutator equation: Limit
  cycle and cooperation},}\ }\href {\doibase 10.1103/PhysRevE.101.042410}
  {\bibfield  {journal} {\bibinfo  {journal} {Phys. Rev. E}\ }\textbf {\bibinfo
  {volume} {101}},\ \bibinfo {pages} {042410} (\bibinfo {year}
  {2020})}\BibitemShut {NoStop}%
\bibitem [{\citenamefont {Szolnoki}\ and\ \citenamefont
  {Perc}(2009)}]{2009_SP_EPL}%
  \BibitemOpen
  \bibfield  {author} {\bibinfo {author} {\bibfnamefont {A.}~\bibnamefont
  {Szolnoki}}\ and\ \bibinfo {author} {\bibfnamefont {M.}~\bibnamefont
  {Perc}},\ }\bibfield  {title} {\enquote {\bibinfo {title} {Resolving social
  dilemmas on evolving random networks},}\ }\href {\doibase
  10.1209/0295-5075/86/30007} {\bibfield  {journal} {\bibinfo  {journal} {EPL}\
  }\textbf {\bibinfo {volume} {86}},\ \bibinfo {pages} {30007} (\bibinfo {year}
  {2009})}\BibitemShut {NoStop}%
\bibitem [{\citenamefont {Amaral}\ \emph {et~al.}(2016)\citenamefont {Amaral},
  \citenamefont {Wardil}, \citenamefont {Perc},\ and\ \citenamefont
  {da~Silva}}]{2016_AWPSJ_PRE}%
  \BibitemOpen
  \bibfield  {author} {\bibinfo {author} {\bibfnamefont {M.~A.}\ \bibnamefont
  {Amaral}}, \bibinfo {author} {\bibfnamefont {L.}~\bibnamefont {Wardil}},
  \bibinfo {author} {\bibfnamefont {M.}~\bibnamefont {Perc}}, \ and\ \bibinfo
  {author} {\bibfnamefont {J.~K.~L.}\ \bibnamefont {da~Silva}},\ }\bibfield
  {title} {\enquote {\bibinfo {title} {Evolutionary mixed games in structured
  populations: Cooperation and the benefits of heterogeneity},}\ }\href
  {\doibase 10.1103/PhysRevE.93.042304} {\bibfield  {journal} {\bibinfo
  {journal} {Phys. Rev. E}\ }\textbf {\bibinfo {volume} {93}},\ \bibinfo
  {pages} {042304} (\bibinfo {year} {2016})}\BibitemShut {NoStop}%
\bibitem [{\citenamefont {Battiston}, \citenamefont {Perc},\ and\ \citenamefont
  {Latora}(2017)}]{2017_BPL_NJP}%
  \BibitemOpen
  \bibfield  {author} {\bibinfo {author} {\bibfnamefont {F.}~\bibnamefont
  {Battiston}}, \bibinfo {author} {\bibfnamefont {M.}~\bibnamefont {Perc}}, \
  and\ \bibinfo {author} {\bibfnamefont {V.}~\bibnamefont {Latora}},\
  }\bibfield  {title} {\enquote {\bibinfo {title} {Determinants of public
  cooperation in multiplex networks},}\ }\href {\doibase
  10.1088/1367-2630/aa6ea1} {\bibfield  {journal} {\bibinfo  {journal} {New J.
  Phys.}\ }\textbf {\bibinfo {volume} {19}},\ \bibinfo {pages} {073017}
  (\bibinfo {year} {2017})}\BibitemShut {NoStop}%
\bibitem [{\citenamefont {Li}\ \emph {et~al.}(2019)\citenamefont {Li},
  \citenamefont {Ma}, \citenamefont {Du},\ and\ \citenamefont
  {Han}}]{2019_L_EPL}%
  \BibitemOpen
  \bibfield  {author} {\bibinfo {author} {\bibfnamefont {D.}~\bibnamefont
  {Li}}, \bibinfo {author} {\bibfnamefont {Z.}~\bibnamefont {Ma}}, \bibinfo
  {author} {\bibfnamefont {J.}~\bibnamefont {Du}}, \ and\ \bibinfo {author}
  {\bibfnamefont {D.}~\bibnamefont {Han}},\ }\bibfield  {title} {\enquote
  {\bibinfo {title} {The evolutionary game with asymmetry and variable
  interaction relations},}\ }\href {\doibase 10.1209/0295-5075/125/10009}
  {\bibfield  {journal} {\bibinfo  {journal} {EPL}\ }\textbf {\bibinfo {volume}
  {125}},\ \bibinfo {pages} {10009} (\bibinfo {year} {2019})}\BibitemShut
  {NoStop}%
\bibitem [{\citenamefont {Taylor}\ and\ \citenamefont
  {Jonker}(1978)}]{1978_TJ_MB}%
  \BibitemOpen
  \bibfield  {author} {\bibinfo {author} {\bibfnamefont {P.~D.}\ \bibnamefont
  {Taylor}}\ and\ \bibinfo {author} {\bibfnamefont {L.~B.}\ \bibnamefont
  {Jonker}},\ }\bibfield  {title} {\enquote {\bibinfo {title} {Evolutionary
  stable strategies and game dynamics},}\ }\href {\doibase
  10.1016/0025-5564(78)90077-9} {\bibfield  {journal} {\bibinfo  {journal}
  {Math. Biosci.}\ }\textbf {\bibinfo {volume} {40}},\ \bibinfo {pages} {145 --
  156} (\bibinfo {year} {1978})}\BibitemShut {NoStop}%
\bibitem [{\citenamefont {Sato}, \citenamefont {Akiyama},\ and\ \citenamefont
  {Farmer}(2002)}]{2002_SAF_PNAS}%
  \BibitemOpen
  \bibfield  {author} {\bibinfo {author} {\bibfnamefont {Y.}~\bibnamefont
  {Sato}}, \bibinfo {author} {\bibfnamefont {E.}~\bibnamefont {Akiyama}}, \
  and\ \bibinfo {author} {\bibfnamefont {J.~D.}\ \bibnamefont {Farmer}},\
  }\bibfield  {title} {\enquote {\bibinfo {title} {Chaos in learning a simple
  two-person game},}\ }\href {\doibase 10.1073/pnas.032086299} {\bibfield
  {journal} {\bibinfo  {journal} {Proc. Natl. Acad. Sci. U.S.A}\ }\textbf
  {\bibinfo {volume} {99}},\ \bibinfo {pages} {4748--4751} (\bibinfo {year}
  {2002})}\BibitemShut {NoStop}%
\bibitem [{\citenamefont {Matsui}(1992)}]{1992_M_JET}%
  \BibitemOpen
  \bibfield  {author} {\bibinfo {author} {\bibfnamefont {A.}~\bibnamefont
  {Matsui}},\ }\bibfield  {title} {\enquote {\bibinfo {title} {Best response
  dynamics and socially stable strategies},}\ }\href {\doibase
  10.1016/0022-0531(92)90040-O} {\bibfield  {journal} {\bibinfo  {journal} {J.
  Econ. Theory}\ }\textbf {\bibinfo {volume} {57}},\ \bibinfo {pages} {343 --
  362} (\bibinfo {year} {1992})}\BibitemShut {NoStop}%
\bibitem [{\citenamefont {Helbing}(1992)}]{1994_H_JMS}%
  \BibitemOpen
  \bibfield  {author} {\bibinfo {author} {\bibfnamefont {D.}~\bibnamefont
  {Helbing}},\ }\enquote {\bibinfo {title} {A mathematical model for behavioral
  changes by pair interactions},}\ in\ \href {\doibase
  10.1007/978-3-642-48808-5_18} {\emph {\bibinfo {booktitle} {Economic
  Evolution and Demographic Change: Formal Models in Social Sciences}}},\
  \bibinfo {editor} {edited by\ \bibinfo {editor} {\bibfnamefont
  {G.}~\bibnamefont {Haag}}, \bibinfo {editor} {\bibfnamefont {U.}~\bibnamefont
  {Mueller}}, \ and\ \bibinfo {editor} {\bibfnamefont {K.~G.}\ \bibnamefont
  {Troitzsch}}}\ (\bibinfo  {publisher} {Springer Berlin Heidelberg},\ \bibinfo
  {address} {Berlin, Heidelberg},\ \bibinfo {year} {1992})\ pp.\ \bibinfo
  {pages} {330--348}\BibitemShut {NoStop}%
\bibitem [{\citenamefont {Sethi}(2000)}]{2000_S_GEB}%
  \BibitemOpen
  \bibfield  {author} {\bibinfo {author} {\bibfnamefont {R.}~\bibnamefont
  {Sethi}},\ }\bibfield  {title} {\enquote {\bibinfo {title} {Stability of
  equilibria in games with procedurally rational players},}\ }\href {\doibase
  https://doi.org/10.1006/game.1999.0753} {\bibfield  {journal} {\bibinfo
  {journal} {Games Econ. Behav.}\ }\textbf {\bibinfo {volume} {32}},\ \bibinfo
  {pages} {85 -- 104} (\bibinfo {year} {2000})}\BibitemShut {NoStop}%
\bibitem [{\citenamefont {Castellano}, \citenamefont {Fortunato},\ and\
  \citenamefont {Loreto}(2009)}]{2009_CFL_RMP}%
  \BibitemOpen
  \bibfield  {author} {\bibinfo {author} {\bibfnamefont {C.}~\bibnamefont
  {Castellano}}, \bibinfo {author} {\bibfnamefont {S.}~\bibnamefont
  {Fortunato}}, \ and\ \bibinfo {author} {\bibfnamefont {V.}~\bibnamefont
  {Loreto}},\ }\bibfield  {title} {\enquote {\bibinfo {title} {Statistical
  physics of social dynamics},}\ }\href {\doibase 10.1103/RevModPhys.81.591}
  {\bibfield  {journal} {\bibinfo  {journal} {Rev. Mod. Phys.}\ }\textbf
  {\bibinfo {volume} {81}},\ \bibinfo {pages} {591--646} (\bibinfo {year}
  {2009})}\BibitemShut {NoStop}%
\bibitem [{\citenamefont {Kuramoto}(1975)}]{1975_Kuramoto}%
  \BibitemOpen
  \bibfield  {author} {\bibinfo {author} {\bibfnamefont {Y.}~\bibnamefont
  {Kuramoto}},\ }\href {\doibase 10.1007/bfb0013294} {\emph {\bibinfo {title}
  {International Symposium on Mathematical Problems in Theoretical Physics}}},\
  edited by\ \bibinfo {editor} {\bibfnamefont {H.}~\bibnamefont {Araki}}\
  (\bibinfo  {publisher} {Springer Berlin Heidelberg},\ \bibinfo {year}
  {1975})\BibitemShut {NoStop}%
\bibitem [{\citenamefont {Antonioni}\ and\ \citenamefont
  {Cardillo}(2017)}]{2017_Antonioni_Cardillo_PRL}%
  \BibitemOpen
  \bibfield  {author} {\bibinfo {author} {\bibfnamefont {A.}~\bibnamefont
  {Antonioni}}\ and\ \bibinfo {author} {\bibfnamefont {A.}~\bibnamefont
  {Cardillo}},\ }\bibfield  {title} {\enquote {\bibinfo {title} {Coevolution of
  synchronization and cooperation in costly networked interactions},}\ }\href
  {\doibase 10.1103/PhysRevLett.118.238301} {\bibfield  {journal} {\bibinfo
  {journal} {Phys. Rev. Lett.}\ }\textbf {\bibinfo {volume} {118}},\ \bibinfo
  {pages} {238301} (\bibinfo {year} {2017})}\BibitemShut {NoStop}%
\bibitem [{\citenamefont {Acebr\'on}\ \emph {et~al.}(2005)\citenamefont
  {Acebr\'on}, \citenamefont {Bonilla}, \citenamefont {P\'erez~Vicente},
  \citenamefont {Ritort},\ and\ \citenamefont {Spigler}}]{2005_ABVCR_RMP}%
  \BibitemOpen
  \bibfield  {author} {\bibinfo {author} {\bibfnamefont {J.~A.}\ \bibnamefont
  {Acebr\'on}}, \bibinfo {author} {\bibfnamefont {L.~L.}\ \bibnamefont
  {Bonilla}}, \bibinfo {author} {\bibfnamefont {C.~J.}\ \bibnamefont
  {P\'erez~Vicente}}, \bibinfo {author} {\bibfnamefont {F.}~\bibnamefont
  {Ritort}}, \ and\ \bibinfo {author} {\bibfnamefont {R.}~\bibnamefont
  {Spigler}},\ }\bibfield  {title} {\enquote {\bibinfo {title} {The kuramoto
  model: A simple paradigm for synchronization phenomena},}\ }\href {\doibase
  10.1103/RevModPhys.77.137} {\bibfield  {journal} {\bibinfo  {journal} {Rev.
  Mod. Phys.}\ }\textbf {\bibinfo {volume} {77}},\ \bibinfo {pages} {137--185}
  (\bibinfo {year} {2005})}\BibitemShut {NoStop}%
\bibitem [{\citenamefont {Boccaletti}\ \emph {et~al.}(2006)\citenamefont
  {Boccaletti}, \citenamefont {Latora}, \citenamefont {Moreno}, \citenamefont
  {Chavez},\ and\ \citenamefont {Hwang}}]{2006_BLMCH_PR}%
  \BibitemOpen
  \bibfield  {author} {\bibinfo {author} {\bibfnamefont {S.}~\bibnamefont
  {Boccaletti}}, \bibinfo {author} {\bibfnamefont {V.}~\bibnamefont {Latora}},
  \bibinfo {author} {\bibfnamefont {Y.}~\bibnamefont {Moreno}}, \bibinfo
  {author} {\bibfnamefont {M.}~\bibnamefont {Chavez}}, \ and\ \bibinfo {author}
  {\bibfnamefont {D.-U.}\ \bibnamefont {Hwang}},\ }\bibfield  {title} {\enquote
  {\bibinfo {title} {Complex networks: Structure and dynamics},}\ }\href
  {\doibase 10.1016/j.physrep.2005.10.009} {\bibfield  {journal} {\bibinfo
  {journal} {Phys. Rep.}\ }\textbf {\bibinfo {volume} {424}},\ \bibinfo {pages}
  {175 -- 308} (\bibinfo {year} {2006})}\BibitemShut {NoStop}%
\bibitem [{\citenamefont {Arenas}\ \emph {et~al.}(2008)\citenamefont {Arenas},
  \citenamefont {Díaz-Guilera}, \citenamefont {Kurths}, \citenamefont
  {Moreno},\ and\ \citenamefont {Zhou}}]{2008_AGKMZ_PR}%
  \BibitemOpen
  \bibfield  {author} {\bibinfo {author} {\bibfnamefont {A.}~\bibnamefont
  {Arenas}}, \bibinfo {author} {\bibfnamefont {A.}~\bibnamefont
  {Díaz-Guilera}}, \bibinfo {author} {\bibfnamefont {J.}~\bibnamefont
  {Kurths}}, \bibinfo {author} {\bibfnamefont {Y.}~\bibnamefont {Moreno}}, \
  and\ \bibinfo {author} {\bibfnamefont {C.}~\bibnamefont {Zhou}},\ }\bibfield
  {title} {\enquote {\bibinfo {title} {Synchronization in complex networks},}\
  }\href {\doibase https://doi.org/10.1016/j.physrep.2008.09.002} {\bibfield
  {journal} {\bibinfo  {journal} {Phys. Rep.}\ }\textbf {\bibinfo {volume}
  {469}},\ \bibinfo {pages} {93 -- 153} (\bibinfo {year} {2008})}\BibitemShut
  {NoStop}%
\bibitem [{\citenamefont {Watts}\ and\ \citenamefont
  {Strogatz}(1998)}]{1998_WS_Nature}%
  \BibitemOpen
  \bibfield  {author} {\bibinfo {author} {\bibfnamefont {D.~J.}\ \bibnamefont
  {Watts}}\ and\ \bibinfo {author} {\bibfnamefont {S.~H.}\ \bibnamefont
  {Strogatz}},\ }\bibfield  {title} {\enquote {\bibinfo {title} {Collective
  dynamics of `small-world' networks},}\ }\href {\doibase 10.1038/30918}
  {\bibfield  {journal} {\bibinfo  {journal} {Nature}\ }\textbf {\bibinfo
  {volume} {393}},\ \bibinfo {pages} {440--442} (\bibinfo {year}
  {1998})}\BibitemShut {NoStop}%
\bibitem [{\citenamefont {Liu}, \citenamefont {Wu},\ and\ \citenamefont
  {Guan}(2018)}]{2018_Liu_Wu_Guan_EPL}%
  \BibitemOpen
  \bibfield  {author} {\bibinfo {author} {\bibfnamefont {X.-S.}\ \bibnamefont
  {Liu}}, \bibinfo {author} {\bibfnamefont {Z.-X.}\ \bibnamefont {Wu}}, \ and\
  \bibinfo {author} {\bibfnamefont {J.-Y.}\ \bibnamefont {Guan}},\ }\bibfield
  {title} {\enquote {\bibinfo {title} {Influence of small-world topology and
  time-scale in evolutionary kuramoto dilemma},}\ }\href {\doibase
  10.1209/0295-5075/122/20001} {\bibfield  {journal} {\bibinfo  {journal}
  {{EPL}}\ }\textbf {\bibinfo {volume} {122}},\ \bibinfo {pages} {20001}
  (\bibinfo {year} {2018})}\BibitemShut {NoStop}%
\bibitem [{\citenamefont {Barab{\'{a}}si}\ and\ \citenamefont
  {Albert}(1999)}]{1999_BA_S}%
  \BibitemOpen
  \bibfield  {author} {\bibinfo {author} {\bibfnamefont {A.-L.}\ \bibnamefont
  {Barab{\'{a}}si}}\ and\ \bibinfo {author} {\bibfnamefont {R.}~\bibnamefont
  {Albert}},\ }\bibfield  {title} {\enquote {\bibinfo {title} {Emergence of
  scaling in random networks},}\ }\href {\doibase 10.1126/science.286.5439.509}
  {\bibfield  {journal} {\bibinfo  {journal} {Science}\ }\textbf {\bibinfo
  {volume} {286}},\ \bibinfo {pages} {509--512} (\bibinfo {year}
  {1999})}\BibitemShut {NoStop}%
\bibitem [{\citenamefont {Erd\"os}\ and\ \citenamefont
  {R\'enyi}(1959)}]{1959_ER_PMD}%
  \BibitemOpen
  \bibfield  {author} {\bibinfo {author} {\bibfnamefont {P.}~\bibnamefont
  {Erd\"os}}\ and\ \bibinfo {author} {\bibfnamefont {A.}~\bibnamefont
  {R\'enyi}},\ }\bibfield  {title} {\enquote {\bibinfo {title} {On random
  graphs i},}\ }\href@noop {} {\bibfield  {journal} {\bibinfo  {journal} {Publ.
  Math. Debr.}\ }\textbf {\bibinfo {volume} {6}},\ \bibinfo {pages} {290--297}
  (\bibinfo {year} {1959})}\BibitemShut {NoStop}%
\bibitem [{\citenamefont {Yang}, \citenamefont {Zhou},\ and\ \citenamefont
  {Wu}(2018)}]{2018_Yang_Han-Xin_Zhou}%
  \BibitemOpen
  \bibfield  {author} {\bibinfo {author} {\bibfnamefont {H.-X.}\ \bibnamefont
  {Yang}}, \bibinfo {author} {\bibfnamefont {T.}~\bibnamefont {Zhou}}, \ and\
  \bibinfo {author} {\bibfnamefont {Z.-X.}\ \bibnamefont {Wu}},\ }\bibfield
  {title} {\enquote {\bibinfo {title} {Kuramoto dilemma alleviated by
  optimizing connectivity and rationality},}\ }\href {\doibase
  10.1103/PhysRevE.98.022201} {\bibfield  {journal} {\bibinfo  {journal} {Phys.
  Rev. E}\ }\textbf {\bibinfo {volume} {98}},\ \bibinfo {pages} {022201}
  (\bibinfo {year} {2018})}\BibitemShut {NoStop}%
\bibitem [{\citenamefont {Watanabe}\ and\ \citenamefont
  {Strogatz}(1993)}]{1993_WS_PRL}%
  \BibitemOpen
  \bibfield  {author} {\bibinfo {author} {\bibfnamefont {S.}~\bibnamefont
  {Watanabe}}\ and\ \bibinfo {author} {\bibfnamefont {S.~H.}\ \bibnamefont
  {Strogatz}},\ }\bibfield  {title} {\enquote {\bibinfo {title} {Integrability
  of a globally coupled oscillator array},}\ }\href {\doibase
  10.1103/PhysRevLett.70.2391} {\bibfield  {journal} {\bibinfo  {journal}
  {Phys. Rev. Lett.}\ }\textbf {\bibinfo {volume} {70}},\ \bibinfo {pages}
  {2391--2394} (\bibinfo {year} {1993})}\BibitemShut {NoStop}%
\bibitem [{\citenamefont {Bick}, \citenamefont {Panaggio},\ and\ \citenamefont
  {Martens}(2018)}]{2018_BPM_C}%
  \BibitemOpen
  \bibfield  {author} {\bibinfo {author} {\bibfnamefont {C.}~\bibnamefont
  {Bick}}, \bibinfo {author} {\bibfnamefont {M.~J.}\ \bibnamefont {Panaggio}},
  \ and\ \bibinfo {author} {\bibfnamefont {E.~A.}\ \bibnamefont {Martens}},\
  }\bibfield  {title} {\enquote {\bibinfo {title} {Chaos in kuramoto oscillator
  networks},}\ }\href {\doibase 10.1063/1.5041444} {\bibfield  {journal}
  {\bibinfo  {journal} {Chaos}\ }\textbf {\bibinfo {volume} {28}},\ \bibinfo
  {pages} {071102} (\bibinfo {year} {2018})}\BibitemShut {NoStop}%
\bibitem [{\citenamefont {May}(1976)}]{1976_May_N}%
  \BibitemOpen
  \bibfield  {author} {\bibinfo {author} {\bibfnamefont {R.~M.}\ \bibnamefont
  {May}},\ }\bibfield  {title} {\enquote {\bibinfo {title} {Simple mathematical
  models with very complicated dynamics},}\ }\href {\doibase 10.1038/261459a0}
  {\bibfield  {journal} {\bibinfo  {journal} {Nature}\ }\textbf {\bibinfo
  {volume} {261}},\ \bibinfo {pages} {459--467} (\bibinfo {year}
  {1976})}\BibitemShut {NoStop}%
\bibitem [{\citenamefont {Kaneko}\ and\ \citenamefont
  {Yanagita}(2014)}]{2014_Kaneko}%
  \BibitemOpen
  \bibfield  {author} {\bibinfo {author} {\bibfnamefont {K.}~\bibnamefont
  {Kaneko}}\ and\ \bibinfo {author} {\bibfnamefont {T.}~\bibnamefont
  {Yanagita}},\ }\bibfield  {title} {\enquote {\bibinfo {title} {{C}oupled
  maps},}\ }\href {\doibase 10.4249/scholarpedia.4085} {\bibfield  {journal}
  {\bibinfo  {journal} {Scholarpedia}\ }\textbf {\bibinfo {volume} {9}},\
  \bibinfo {pages} {4085} (\bibinfo {year} {2014})},\ \bibinfo {note} {revision
  \#149460}\BibitemShut {NoStop}%
\bibitem [{\citenamefont {Kaneko}(1993)}]{1993_Kaneko}%
  \BibitemOpen
  \bibfield  {author} {\bibinfo {author} {\bibfnamefont {K.}~\bibnamefont
  {Kaneko}},\ }\href@noop {} {\emph {\bibinfo {title} {Theory and applications
  of coupled map lattices}}}\ (\bibinfo  {publisher} {Chichester: Wiley, New
  York},\ \bibinfo {year} {1993})\BibitemShut {NoStop}%
\bibitem [{\citenamefont {Jost}\ and\ \citenamefont {Joy}(2001)}]{2001_JJ_PRE}%
  \BibitemOpen
  \bibfield  {author} {\bibinfo {author} {\bibfnamefont {J.}~\bibnamefont
  {Jost}}\ and\ \bibinfo {author} {\bibfnamefont {M.~P.}\ \bibnamefont {Joy}},\
  }\bibfield  {title} {\enquote {\bibinfo {title} {Spectral properties and
  synchronization in coupled map lattices},}\ }\href {\doibase
  10.1103/PhysRevE.65.016201} {\bibfield  {journal} {\bibinfo  {journal} {Phys.
  Rev. E}\ }\textbf {\bibinfo {volume} {65}},\ \bibinfo {pages} {016201}
  (\bibinfo {year} {2001})}\BibitemShut {NoStop}%
\bibitem [{\citenamefont {Mart\'{\i}}\ and\ \citenamefont
  {Masoller}(2003)}]{2003_MM_PRE}%
  \BibitemOpen
  \bibfield  {author} {\bibinfo {author} {\bibfnamefont {A.~C.}\ \bibnamefont
  {Mart\'{\i}}}\ and\ \bibinfo {author} {\bibfnamefont {C.}~\bibnamefont
  {Masoller}},\ }\bibfield  {title} {\enquote {\bibinfo {title} {Delay-induced
  synchronization phenomena in an array of globally coupled logistic maps},}\
  }\href {\doibase 10.1103/PhysRevE.67.056219} {\bibfield  {journal} {\bibinfo
  {journal} {Phys. Rev. E}\ }\textbf {\bibinfo {volume} {67}},\ \bibinfo
  {pages} {056219} (\bibinfo {year} {2003})}\BibitemShut {NoStop}%
\bibitem [{\citenamefont {Lind}, \citenamefont {Gallas},\ and\ \citenamefont
  {Herrmann}(2004)}]{2004_LGH_PRE}%
  \BibitemOpen
  \bibfield  {author} {\bibinfo {author} {\bibfnamefont {P.~G.}\ \bibnamefont
  {Lind}}, \bibinfo {author} {\bibfnamefont {J.~A.~C.}\ \bibnamefont {Gallas}},
  \ and\ \bibinfo {author} {\bibfnamefont {H.~J.}\ \bibnamefont {Herrmann}},\
  }\bibfield  {title} {\enquote {\bibinfo {title} {Coherence in scale-free
  networks of chaotic maps},}\ }\href {\doibase 10.1103/PhysRevE.70.056207}
  {\bibfield  {journal} {\bibinfo  {journal} {Phys. Rev. E}\ }\textbf {\bibinfo
  {volume} {70}},\ \bibinfo {pages} {056207} (\bibinfo {year}
  {2004})}\BibitemShut {NoStop}%
\bibitem [{\citenamefont {Jalan}, \citenamefont {Amritkar},\ and\ \citenamefont
  {Hu}(2005)}]{2005_JAH_PRE}%
  \BibitemOpen
  \bibfield  {author} {\bibinfo {author} {\bibfnamefont {S.}~\bibnamefont
  {Jalan}}, \bibinfo {author} {\bibfnamefont {R.~E.}\ \bibnamefont {Amritkar}},
  \ and\ \bibinfo {author} {\bibfnamefont {C.-K.}\ \bibnamefont {Hu}},\
  }\bibfield  {title} {\enquote {\bibinfo {title} {Synchronized clusters in
  coupled map networks. i. numerical studies},}\ }\href {\doibase
  10.1103/PhysRevE.72.016211} {\bibfield  {journal} {\bibinfo  {journal} {Phys.
  Rev. E}\ }\textbf {\bibinfo {volume} {72}},\ \bibinfo {pages} {016211}
  (\bibinfo {year} {2005})}\BibitemShut {NoStop}%
\bibitem [{\citenamefont {Amritkar}, \citenamefont {Jalan},\ and\ \citenamefont
  {Hu}(2005)}]{2005_AJH_PRE}%
  \BibitemOpen
  \bibfield  {author} {\bibinfo {author} {\bibfnamefont {R.~E.}\ \bibnamefont
  {Amritkar}}, \bibinfo {author} {\bibfnamefont {S.}~\bibnamefont {Jalan}}, \
  and\ \bibinfo {author} {\bibfnamefont {C.-K.}\ \bibnamefont {Hu}},\
  }\bibfield  {title} {\enquote {\bibinfo {title} {Synchronized clusters in
  coupled map networks. ii. stability analysis},}\ }\href {\doibase
  10.1103/PhysRevE.72.016212} {\bibfield  {journal} {\bibinfo  {journal} {Phys.
  Rev. E}\ }\textbf {\bibinfo {volume} {72}},\ \bibinfo {pages} {016212}
  (\bibinfo {year} {2005})}\BibitemShut {NoStop}%
\bibitem [{\citenamefont {C.}, \citenamefont {Masoller},\ and\ \citenamefont
  {Mart{\'{\i}}}(2008)}]{2008_PMM_EPJB}%
  \BibitemOpen
  \bibfield  {author} {\bibinfo {author} {\bibfnamefont {M.~P.}\ \bibnamefont
  {C.}}, \bibinfo {author} {\bibfnamefont {C.}~\bibnamefont {Masoller}}, \ and\
  \bibinfo {author} {\bibfnamefont {A.~C.}\ \bibnamefont {Mart{\'{\i}}}},\
  }\bibfield  {title} {\enquote {\bibinfo {title} {Synchronizability of chaotic
  logistic maps in delayed complex networks},}\ }\href {\doibase
  10.1140/epjb/e2008-00467-3} {\bibfield  {journal} {\bibinfo  {journal} {Eur.
  Phys. J. B}\ }\textbf {\bibinfo {volume} {67}},\ \bibinfo {pages} {83--93}
  (\bibinfo {year} {2008})}\BibitemShut {NoStop}%
\bibitem [{\citenamefont {Li}\ \emph {et~al.}(2009)\citenamefont {Li},
  \citenamefont {Xu}, \citenamefont {Sun}, \citenamefont {Xu},\ and\
  \citenamefont {Kurths}}]{2009_LXSXK_Chaos}%
  \BibitemOpen
  \bibfield  {author} {\bibinfo {author} {\bibfnamefont {C.}~\bibnamefont
  {Li}}, \bibinfo {author} {\bibfnamefont {C.}~\bibnamefont {Xu}}, \bibinfo
  {author} {\bibfnamefont {W.}~\bibnamefont {Sun}}, \bibinfo {author}
  {\bibfnamefont {J.}~\bibnamefont {Xu}}, \ and\ \bibinfo {author}
  {\bibfnamefont {J.}~\bibnamefont {Kurths}},\ }\bibfield  {title} {\enquote
  {\bibinfo {title} {Outer synchronization of coupled discrete-time
  networks},}\ }\href {\doibase 10.1063/1.3068357} {\bibfield  {journal}
  {\bibinfo  {journal} {Chaos}\ }\textbf {\bibinfo {volume} {19}},\ \bibinfo
  {pages} {013106} (\bibinfo {year} {2009})}\BibitemShut {NoStop}%
\bibitem [{\citenamefont {Mirollo}\ and\ \citenamefont
  {Strogatz}(1990)}]{1990_MS}%
  \BibitemOpen
  \bibfield  {author} {\bibinfo {author} {\bibfnamefont {R.~E.}\ \bibnamefont
  {Mirollo}}\ and\ \bibinfo {author} {\bibfnamefont {S.~H.}\ \bibnamefont
  {Strogatz}},\ }\bibfield  {title} {\enquote {\bibinfo {title} {Amplitude
  death in an array of limit-cycle oscillators},}\ }\href {\doibase
  10.1007/bf01013676} {\bibfield  {journal} {\bibinfo  {journal} {J. Stat.
  Phys.}\ }\textbf {\bibinfo {volume} {60}},\ \bibinfo {pages} {245--262}
  (\bibinfo {year} {1990})}\BibitemShut {NoStop}%
\bibitem [{\citenamefont {Ramana~Reddy}, \citenamefont {Sen},\ and\
  \citenamefont {Johnston}(1998)}]{1998_RSJ}%
  \BibitemOpen
  \bibfield  {author} {\bibinfo {author} {\bibfnamefont {D.~V.}\ \bibnamefont
  {Ramana~Reddy}}, \bibinfo {author} {\bibfnamefont {A.}~\bibnamefont {Sen}}, \
  and\ \bibinfo {author} {\bibfnamefont {G.~L.}\ \bibnamefont {Johnston}},\
  }\bibfield  {title} {\enquote {\bibinfo {title} {Time delay induced death in
  coupled limit cycle oscillators},}\ }\href {\doibase
  10.1103/PhysRevLett.80.5109} {\bibfield  {journal} {\bibinfo  {journal}
  {Phys. Rev. Lett.}\ }\textbf {\bibinfo {volume} {80}},\ \bibinfo {pages}
  {5109--5112} (\bibinfo {year} {1998})}\BibitemShut {NoStop}%
\bibitem [{\citenamefont {Sinha}(2002)}]{2002_Sinha_PRE}%
  \BibitemOpen
  \bibfield  {author} {\bibinfo {author} {\bibfnamefont {S.}~\bibnamefont
  {Sinha}},\ }\bibfield  {title} {\enquote {\bibinfo {title} {Random coupling
  of chaotic maps leads to spatiotemporal synchronization},}\ }\href {\doibase
  10.1103/PhysRevE.66.016209} {\bibfield  {journal} {\bibinfo  {journal} {Phys.
  Rev. E}\ }\textbf {\bibinfo {volume} {66}},\ \bibinfo {pages} {016209}
  (\bibinfo {year} {2002})}\BibitemShut {NoStop}%
\bibitem [{\citenamefont {Wang}\ \emph {et~al.}(2016)\citenamefont {Wang},
  \citenamefont {Du}, \citenamefont {Jin}, \citenamefont {Wu},\ and\
  \citenamefont {Qu}}]{2016_Wang}%
  \BibitemOpen
  \bibfield  {author} {\bibinfo {author} {\bibfnamefont {S.-J.}\ \bibnamefont
  {Wang}}, \bibinfo {author} {\bibfnamefont {R.-H.}\ \bibnamefont {Du}},
  \bibinfo {author} {\bibfnamefont {T.}~\bibnamefont {Jin}}, \bibinfo {author}
  {\bibfnamefont {X.-S.}\ \bibnamefont {Wu}}, \ and\ \bibinfo {author}
  {\bibfnamefont {S.-X.}\ \bibnamefont {Qu}},\ }\bibfield  {title} {\enquote
  {\bibinfo {title} {Synchronous slowing down in coupled logistic maps via
  random network topology},}\ }\href {\doibase 10.1038/srep23448} {\bibfield
  {journal} {\bibinfo  {journal} {Sci. Rep.}\ }\textbf {\bibinfo {volume} {6}}
  (\bibinfo {year} {2016}),\ 10.1038/srep23448}\BibitemShut {NoStop}%
\bibitem [{\citenamefont {Blume}(1993)}]{1993_B_GEB}%
  \BibitemOpen
  \bibfield  {author} {\bibinfo {author} {\bibfnamefont {L.~E.}\ \bibnamefont
  {Blume}},\ }\bibfield  {title} {\enquote {\bibinfo {title} {The statistical
  mechanics of strategic interaction},}\ }\href {\doibase
  https://doi.org/10.1006/game.1993.1023} {\bibfield  {journal} {\bibinfo
  {journal} {Games Econ. Behav.}\ }\textbf {\bibinfo {volume} {5}},\ \bibinfo
  {pages} {387 -- 424} (\bibinfo {year} {1993})}\BibitemShut {NoStop}%
\bibitem [{\citenamefont {Szab\'o}\ and\ \citenamefont
  {T\ifmmode~\mbox{\H{o}}\else \H{o}\fi{}ke}(1998)}]{1998_SHH_PRE}%
  \BibitemOpen
  \bibfield  {author} {\bibinfo {author} {\bibfnamefont {G.}~\bibnamefont
  {Szab\'o}}\ and\ \bibinfo {author} {\bibfnamefont {C.}~\bibnamefont
  {T\ifmmode~\mbox{\H{o}}\else \H{o}\fi{}ke}},\ }\bibfield  {title} {\enquote
  {\bibinfo {title} {Evolutionary prisoner's dilemma game on a square
  lattice},}\ }\href {\doibase 10.1103/PhysRevE.58.69} {\bibfield  {journal}
  {\bibinfo  {journal} {Phys. Rev. E}\ }\textbf {\bibinfo {volume} {58}},\
  \bibinfo {pages} {69--73} (\bibinfo {year} {1998})}\BibitemShut {NoStop}%
\bibitem [{\citenamefont {Roca}, \citenamefont {Cuesta},\ and\ \citenamefont
  {Sánchez}(2009)}]{2009_RCS_PLR}%
  \BibitemOpen
  \bibfield  {author} {\bibinfo {author} {\bibfnamefont {C.~P.}\ \bibnamefont
  {Roca}}, \bibinfo {author} {\bibfnamefont {J.~A.}\ \bibnamefont {Cuesta}}, \
  and\ \bibinfo {author} {\bibfnamefont {A.}~\bibnamefont {Sánchez}},\
  }\bibfield  {title} {\enquote {\bibinfo {title} {Evolutionary game theory:
  Temporal and spatial effects beyond replicator dynamics},}\ }\href {\doibase
  10.1016/j.plrev.2009.08.001} {\bibfield  {journal} {\bibinfo  {journal}
  {Phys. Life Rev.}\ }\textbf {\bibinfo {volume} {6}},\ \bibinfo {pages} {208
  -- 249} (\bibinfo {year} {2009})}\BibitemShut {NoStop}%
\bibitem [{\citenamefont {Hagberg}\ and\ \citenamefont
  {Schult}(2008)}]{2008_HS_C}%
  \BibitemOpen
  \bibfield  {author} {\bibinfo {author} {\bibfnamefont {A.}~\bibnamefont
  {Hagberg}}\ and\ \bibinfo {author} {\bibfnamefont {D.~A.}\ \bibnamefont
  {Schult}},\ }\bibfield  {title} {\enquote {\bibinfo {title} {Rewiring
  networks for synchronization},}\ }\href {\doibase 10.1063/1.2975842}
  {\bibfield  {journal} {\bibinfo  {journal} {Chaos}\ }\textbf {\bibinfo
  {volume} {18}},\ \bibinfo {pages} {037105} (\bibinfo {year}
  {2008})}\BibitemShut {NoStop}%
\bibitem [{\citenamefont {Chen}, \citenamefont {Qiu},\ and\ \citenamefont
  {Huang}(2009)}]{2009_CQH_PRE}%
  \BibitemOpen
  \bibfield  {author} {\bibinfo {author} {\bibfnamefont {L.}~\bibnamefont
  {Chen}}, \bibinfo {author} {\bibfnamefont {C.}~\bibnamefont {Qiu}}, \ and\
  \bibinfo {author} {\bibfnamefont {H.~B.}\ \bibnamefont {Huang}},\ }\bibfield
  {title} {\enquote {\bibinfo {title} {Synchronization with on-off coupling:
  Role of time scales in network dynamics},}\ }\href {\doibase
  10.1103/PhysRevE.79.045101} {\bibfield  {journal} {\bibinfo  {journal} {Phys.
  Rev. E}\ }\textbf {\bibinfo {volume} {79}},\ \bibinfo {pages} {045101}
  (\bibinfo {year} {2009})}\BibitemShut {NoStop}%
\bibitem [{\citenamefont {Schr\"oder}\ \emph {et~al.}(2015)\citenamefont
  {Schr\"oder}, \citenamefont {Mannattil}, \citenamefont {Dutta}, \citenamefont
  {Chakraborty},\ and\ \citenamefont {Timme}}]{2015_SMDCT_PRL}%
  \BibitemOpen
  \bibfield  {author} {\bibinfo {author} {\bibfnamefont {M.}~\bibnamefont
  {Schr\"oder}}, \bibinfo {author} {\bibfnamefont {M.}~\bibnamefont
  {Mannattil}}, \bibinfo {author} {\bibfnamefont {D.}~\bibnamefont {Dutta}},
  \bibinfo {author} {\bibfnamefont {S.}~\bibnamefont {Chakraborty}}, \ and\
  \bibinfo {author} {\bibfnamefont {M.}~\bibnamefont {Timme}},\ }\bibfield
  {title} {\enquote {\bibinfo {title} {Transient uncoupling induces
  synchronization},}\ }\href {\doibase 10.1103/PhysRevLett.115.054101}
  {\bibfield  {journal} {\bibinfo  {journal} {Phys. Rev. Lett.}\ }\textbf
  {\bibinfo {volume} {115}},\ \bibinfo {pages} {054101} (\bibinfo {year}
  {2015})}\BibitemShut {NoStop}%
\bibitem [{\citenamefont {Zhou}\ \emph {et~al.}(2016)\citenamefont {Zhou},
  \citenamefont {Zou}, \citenamefont {Guan}, \citenamefont {Liu},\ and\
  \citenamefont {Boccaletti}}]{2016_ZZGLB_SR}%
  \BibitemOpen
  \bibfield  {author} {\bibinfo {author} {\bibfnamefont {J.}~\bibnamefont
  {Zhou}}, \bibinfo {author} {\bibfnamefont {Y.}~\bibnamefont {Zou}}, \bibinfo
  {author} {\bibfnamefont {S.}~\bibnamefont {Guan}}, \bibinfo {author}
  {\bibfnamefont {Z.}~\bibnamefont {Liu}}, \ and\ \bibinfo {author}
  {\bibfnamefont {S.}~\bibnamefont {Boccaletti}},\ }\bibfield  {title}
  {\enquote {\bibinfo {title} {Synchronization in slowly switching networks of
  coupled oscillators},}\ }\href {\doibase 10.1038/srep35979} {\bibfield
  {journal} {\bibinfo  {journal} {Sci. Rep.}\ }\textbf {\bibinfo {volume} {6}}
  (\bibinfo {year} {2016}),\ 10.1038/srep35979}\BibitemShut {NoStop}%
\bibitem [{\citenamefont {Tandon}\ \emph {et~al.}(2016)\citenamefont {Tandon},
  \citenamefont {Schr\"{o}der}, \citenamefont {Mannattil}, \citenamefont
  {Timme},\ and\ \citenamefont {Chakraborty}}]{2016_TSMTS_Chaos}%
  \BibitemOpen
  \bibfield  {author} {\bibinfo {author} {\bibfnamefont {A.}~\bibnamefont
  {Tandon}}, \bibinfo {author} {\bibfnamefont {M.}~\bibnamefont
  {Schr\"{o}der}}, \bibinfo {author} {\bibfnamefont {M.}~\bibnamefont
  {Mannattil}}, \bibinfo {author} {\bibfnamefont {M.}~\bibnamefont {Timme}}, \
  and\ \bibinfo {author} {\bibfnamefont {S.}~\bibnamefont {Chakraborty}},\
  }\bibfield  {title} {\enquote {\bibinfo {title} {Synchronizing noisy
  nonidentical oscillators by transient uncoupling},}\ }\href {\doibase
  10.1063/1.4959141} {\bibfield  {journal} {\bibinfo  {journal} {Chaos}\
  }\textbf {\bibinfo {volume} {26}},\ \bibinfo {pages} {094817} (\bibinfo
  {year} {2016})}\BibitemShut {NoStop}%
\bibitem [{\citenamefont {Schr\"{o}der}\ \emph {et~al.}()\citenamefont
  {Schr\"{o}der}, \citenamefont {Chakraborty}, \citenamefont {Witthaut},
  \citenamefont {Nagler},\ and\ \citenamefont {Timme}}]{2016_SCWNT_SR}%
  \BibitemOpen
  \bibfield  {author} {\bibinfo {author} {\bibfnamefont {M.}~\bibnamefont
  {Schr\"{o}der}}, \bibinfo {author} {\bibfnamefont {S.}~\bibnamefont
  {Chakraborty}}, \bibinfo {author} {\bibfnamefont {D.}~\bibnamefont
  {Witthaut}}, \bibinfo {author} {\bibfnamefont {J.}~\bibnamefont {Nagler}}, \
  and\ \bibinfo {author} {\bibfnamefont {M.}~\bibnamefont {Timme}},\ }\bibfield
   {title} {\enquote {\bibinfo {title} {Interaction control to synchronize
  non-synchronizable networks},}\ }\href {\doibase 10.1038/srep37142}
  {\bibfield  {journal} {\bibinfo  {journal} {Sci. Rep.}\ }\textbf {\bibinfo
  {volume} {6}},\ 10.1038/srep37142}\BibitemShut {NoStop}%
\bibitem [{\citenamefont {Li}\ \emph {et~al.}(2018)\citenamefont {Li},
  \citenamefont {Sun}, \citenamefont {Chen},\ and\ \citenamefont
  {Wang}}]{2018_LSCW_PRE}%
  \BibitemOpen
  \bibfield  {author} {\bibinfo {author} {\bibfnamefont {S.}~\bibnamefont
  {Li}}, \bibinfo {author} {\bibfnamefont {N.}~\bibnamefont {Sun}}, \bibinfo
  {author} {\bibfnamefont {L.}~\bibnamefont {Chen}}, \ and\ \bibinfo {author}
  {\bibfnamefont {X.}~\bibnamefont {Wang}},\ }\bibfield  {title} {\enquote
  {\bibinfo {title} {Network synchronization with periodic coupling},}\ }\href
  {\doibase 10.1103/physreve.98.012304} {\bibfield  {journal} {\bibinfo
  {journal} {Phys. Rev. E}\ }\textbf {\bibinfo {volume} {98}} (\bibinfo {year}
  {2018}),\ 10.1103/physreve.98.012304}\BibitemShut {NoStop}%
\bibitem [{\citenamefont {Pandit}, \citenamefont {Mukhopadhyay},\ and\
  \citenamefont {Chakraborty}(2018)}]{2018_PMS_C}%
  \BibitemOpen
  \bibfield  {author} {\bibinfo {author} {\bibfnamefont {V.}~\bibnamefont
  {Pandit}}, \bibinfo {author} {\bibfnamefont {A.}~\bibnamefont
  {Mukhopadhyay}}, \ and\ \bibinfo {author} {\bibfnamefont {S.}~\bibnamefont
  {Chakraborty}},\ }\bibfield  {title} {\enquote {\bibinfo {title} {Weight of
  fitness deviation governs strict physical chaos in replicator dynamics},}\
  }\href {\doibase 10.1063/1.5011955} {\bibfield  {journal} {\bibinfo
  {journal} {Chaos}\ }\textbf {\bibinfo {volume} {28}},\ \bibinfo {pages}
  {033104} (\bibinfo {year} {2018})}\BibitemShut {NoStop}%
\bibitem [{\citenamefont {Mukhopadhyay}\ and\ \citenamefont
  {Chakraborty}(2020)}]{2020_MC_JTB}%
  \BibitemOpen
  \bibfield  {author} {\bibinfo {author} {\bibfnamefont {A.}~\bibnamefont
  {Mukhopadhyay}}\ and\ \bibinfo {author} {\bibfnamefont {S.}~\bibnamefont
  {Chakraborty}},\ }\bibfield  {title} {\enquote {\bibinfo {title} {Periodic
  orbit can be evolutionarily stable: Case study of discrete replicator
  dynamics},}\ }\href {\doibase https://doi.org/10.1016/j.jtbi.2020.110288}
  {\bibfield  {journal} {\bibinfo  {journal} {J. Theor. Biol.}\ }\textbf
  {\bibinfo {volume} {497}},\ \bibinfo {pages} {110288} (\bibinfo {year}
  {2020})}\BibitemShut {NoStop}%
\bibitem [{\citenamefont {McElreath}\ and\ \citenamefont
  {Boyd}(2007)}]{2007_MB}%
  \BibitemOpen
  \bibfield  {author} {\bibinfo {author} {\bibfnamefont {R.}~\bibnamefont
  {McElreath}}\ and\ \bibinfo {author} {\bibfnamefont {R.}~\bibnamefont
  {Boyd}},\ }\href {\doibase 10.7208/chicago/9780226558288.001.0001} {\emph
  {\bibinfo {title} {Mathematical Models of Social Evolution}}}\ (\bibinfo
  {publisher} {University of Chicago Press, Chicago},\ \bibinfo {year}
  {2007})\BibitemShut {NoStop}%
\end{thebibliography}%
 \end{document}